\documentclass[copyright]{eptcs}
\providecommand{\event}{AMCA-POP 2010} % Name of the event you are submitting to

\providecommand{\firstpage}{1}

\usepackage{breakurl}        % Not needed if you use pdflatex only.
\usepackage{graphics}
\usepackage{graphicx}
\usepackage{epsfig}
\usepackage{amssymb}
\usepackage{amsthm}

\newcommand{\virg}{``}
\newcommand{\ra}{\rightarrow}

\title{An Analysis on the Influence of Network Topologies on\\Local and Global Dynamics of Metapopulation Systems}

\author{Daniela Besozzi$^a$ \quad Paolo Cazzaniga$^b$  \\Dario Pescini$^b$  \quad Giancarlo Mauri$^b$\\
\institute{$^a$Universit\`{a} degli Studi di Milano\\ Dipartimento di Informatica e Comunicazione\\ Via Comelico 39, 20135 Milano, Italy}
\email{besozzi@dico.unimi.it}\\
\institute{$^b$Universit\`{a} degli Studi di Milano-Bicocca\\ Dipartimento di Informatica, Sistemistica e Comunicazione\\ Viale Sarca 336, 20126 Milano, Italy}
\email{cazzaniga/pescini/mauri@disco.unimib.it}
}

\def\titlerunning{The influence of network topologies on metapopulations dynamics}
\def\authorrunning{D. Besozzi et al.}
\begin{document}
\maketitle

\begin{abstract}
Metapopulations are models of ecological systems, describing the interactions
and the behavior of populations that live in fragmented habitats. In this paper, we present a model of metapopulations based on the multivolume simulation algorithm tau-DPP, a stochastic class of membrane systems, that we utilize to investigate the influence that different habitat topologies can have on the local and global dynamics of metapopulations. In particular, we focus our analysis on the migration rate of individuals among adjacent patches, and on their capability of colonizing the empty patches in the habitat. We compare the simulation results obtained for each habitat topology, and conclude the paper with some proposals for other research issues concerning metapopulations.
\end{abstract}

\section{Introduction}

The field of metapopulations ecology deals with the study of spatial systems describing the behavior of interacting populations that live in fragmented habitats \cite{Hanski}. The purpose of these models is to understand how the local and global dynamics of metapopulation systems, usually balanced between local extinctions and new colonizations of unoccupied patches, depend on the spatial arrangement of the habitat. Consequently, relevant insights into related fields of ecological research, such as evolutionary ecology or conservation and landscape management, can be achieved. Indeed, the topology of fragmented habitats potentially holds relevant implications for the persistence of populations, and their robustness against natural or anthropogenic disturbance \cite{habitat_mosaics}.

Recently, in addition to ever increasing applications of graph-based methods for the analysis of complex networks in cell biology \cite{graph_cellbiology,scalefree_cellbiology}, graph theory has also been applied to the study of metapopulations systems. In graph models of metapopulations, nodes are used to represent habitat patches, and graph edges are used to denote some functional connections between patches (typically related to the dispersal of individuals). Attributes can be associated to nodes, describing the quality or dimension of patches, while different types of edges can be exploited to represent the distance between connected patches, the rate of dispersal between a couple of patches, or simply whether two patches are connected or not.

Metapopulation models using graph-based methods \cite{habitat_mosaics,spatial_graphs} are simple to implement and require relatively few data for their definition, while individual-based models implement more detailed aspects, concerning the nature and the interaction of populations \cite{spatial_explicit,spatially_explicit_review}. Both types of modeling approaches are useful for the analysis of specific features of metapopulations but, while the first focuses on the properties of the habitat topology, the second is more concerned with the emergent dynamics. In this paper, we present a stochastic multivolume model of metapopulations, which integrates the explicit representation of  interactions between the individuals of the populations -- and therefore allows to simulate the emergent local and global dynamics -- with a graph description of the habitat topology -- which allows to investigate the influence of distinct spatial structures on the dynamics.

This model, which represents a simplified extension of a previous metapopulation model that we introduced in \cite{metapop,bicta07}, is based on the multivolume stochastic simulation algorithm tau-DPP \cite{tauWMC7,VolumeRozenberg}, a stochastic class of membrane systems. Membrane systems, or P systems, were introduced in \cite{Paun00} as a class of unconventional computing devices of distributed, parallel and nondeterministic
type, inspired by the compartmental structure and the functioning of living
cells. The basic model consists of a membrane structure where multisets of
objects evolve according to given evolution rules. A
comprehensive overview of P systems and of its many applications in various research areas, ranging
from Biology to Linguistics to Computer Science, can be found in \cite{PaunBook,VAPS,oxford_handbook_MC}.

In tau-DPP, the distinct compartments of any multivolume model can be arranged according to a specified hierarchy (e.g., a membrane structure), under the additional assumption that the topological structure and the volume dimensions do not change during the system evolution (each volume is assumed to satisfy the standard requirements of the classical stochastic simulation algorithm, see \cite{Gill77} and \cite{BioSimWare} for more details). Inside each volume, two different types of rules can be defined: the \emph{internal rules}, which modify the objects contained inside the volume where they take place (in the case of metapopulation, they describe the growth and death of population individuals according to the Lotka-Volterra model of preys and predators), and the \emph{communication rules}, which are used to move the objects between adjacent volumes (in the case of metapopulation, they describe the migration of population individuals).

In this paper, tau-DPP is exploited to analyze the emergent dynamics of metapopulation systems, where the focus is on the influence that the topology of patches has on the migration of individuals, and their capability to colonize other patches in the habitat. To this purpose, we consider six different habitat topologies, formally described by graph structures, and analyze how the topological structure of patch-to-patch connections, and the rate of individual dispersal between connected patches, influence the local and global dynamics of a metapopulation. In particular, we will first consider how a given topology and a fixed dispersal rate between patches can influence the prey-predators dynamics, and then we will focus on the colonization of empty patches, starting from the dispersal of predators that live in a few patches which occupy peculiar positions in the given network topology.

The paper is structured as follows: in Section \ref{sec:metapop} we present the concept of metapopulations in Ecology, and then describe the multivolume model of metapopulations by focusing, in particular, to the different habitat topologies. In Section \ref{sec:sim_dyn} we will show the simulation results concerning the influence of these habitat topologies on the emergent dynamics of metapopulations, considering the effects of predators dispersal and colonization. Finally, in Section \ref{sec:concl} we conclude the paper with some final remarks and several proposals for further research issues concerning metapopulations.

\section{Metapopulations}\label{sec:metapop}

In this section, we first provide a brief introduction to the most relevant features of metapopulations, concerning both the topology of the habitats and the emergent dynamics. Then, we describe the modeling approach used in this paper, that is based on a stochastic class of membrane systems, which will be used in Section \ref{sec:sim_dyn} to analyze the influence of different network topologies on the dynamics of metapopulations.

\subsection{Dynamical models of interacting populations in Ecology}\label{subsec:eco_metapop}

Since its introduction in \cite{Levins}, the concept of metapopulations (also called
\emph{multi-patch systems}) has been extensively applied in Ecology to analyze the behavior of interacting
populations, to the purpose of determining how fragmented habitats can
influence various aspects of these systems, such as local and global population persistence, or the evolution
of species \cite{genetic}. Lately, this topic has been largely employed for other populations species, living in both
natural and artificial/theoretical fragmented landscapes \cite{Hanski}.

A metapopulation consists of local populations, living in spatially separated
habitats called \emph{patches} -- which can be characterized by different areas, quality
or isolation -- connected each other through a \emph{dispersal pool}, which is the spatial place where
individuals from a population spend some lifetime during the migration among
patches. In multi-patch systems, two principal
types of dynamics exist: on the one hand, the individuals of the different populations can have \emph{local} interactions inside each patch (according to a given dynamical model, e.g., the Lotka-Volterra system of interaction between preys and predators \cite{Murray}); on the other hand, the dispersal of individuals among mutually connected patches can influence the \emph{global} behavior of the whole system \cite{Jansen-2,Jansen,Taylor,Weisser}.
The dispersal of individuals, which is usually dependent on the distance between patches, may reduce the
local population growth, and thus increase the extinction risk, which can
be due also to environmental and demographical stochasticity. Hence, the
persistence of populations is assumed to be balanced between local extinctions
and the process of colonization, that is, the establishment of new populations in empty patches \cite{Hanski}.

Several theoretical frameworks for metapopulation analysis have been defined up
to now, remarking specific properties of multi-patch systems which have been either
explicitly or implicitly considered in these modeling methods (see, e.g., \cite{dunning,Hanski,SPOMSIM,Hastings} for
further details). For instance, referring to the landscape, most theoretical
models take care of the spatial structure of the habitat, the local quality of the environment, the
patch areas and their mutual connectivity (or isolation), in order to capture the effect of
habitat fragmentation on species persistence. In fact, good local conditions can
determine the growth and the survival of populations inside the patches, and high patch
connectivity can decrease local extinction risk. Moreover, as dispersal and
colonization are distance-dependent elements, they can be used to account for the
importance of real landscape structures. Referring to population interactions
and dynamics, colonization can depend or not on the cooperation of migrating
individuals (in the first case, it is called ``Allee effect''). Models not accounting for within-patch dynamics -- but only assuming whether a patch is occupied or not -- usually consider local dynamics
on a faster time scale with respect to the global dynamics, and also neglect the
dependence of colonization and extinction rates on population sizes. Finally,
regional stochasticity can account for ``bad'' or ``good'' years over the local
environmental quality, which depends on, e.g., the weather conditions which
affect sustenance resource availability and, once more, they can influence the
growth and survival of populations.

Recently, graph-based models for metapopulations have started to be more and more defined because of the intuitive and visual way they hold for the representation of these ecological systems (see \cite{habitat_mosaics,MinorUrban2008,Urban_Ecology_2001} and references therein). In these models, nodes represent habitat patches and graph edges denote functional connections between patches (typically related to the dispersal of individuals). In addition, attributes can be associated to nodes, describing the quality or dimension of patches, and different types of edges can be adopted to represent the distance between connected patches, the rate of dispersal between a couple of patches, or simply whether two patches are connected or not. These models allow to make insights into the features of habitat distribution, such as the predominant importance of some nodes or clusters of nodes with respect to other characteristics of metapopulation, like their dynamics, the vulnerability to disturbance, the persistence of populations according to dispersal, and so on. These results open promising perspective in related  research fields as evolutionary ecology, conservation biology, epidemiology, management and design of natural reserves.

\subsection{A P system--based model of metapopulations: focusing on network topologies}\label{subsec:Psyst_metapop}

Most of the issues discussed in Section \ref{subsec:eco_metapop} were explicitly considered in our previous model for metapopulations \cite{bicta07,metapop}. In those works, metapopulation models were based on a class of membrane systems called DPP \cite{DNA11,IJFCS}, which were used to execute qualitative stochastic simulations of the local and global dynamics of metapopulations. In particular, in \cite{metapop} we introduced a model of metapopulations with predator-prey dynamics, where additional features were used in order to catch and better describe relevant properties of the modeled system.
For instance, the regions of the membrane structure were represented as nodes of a weighted graph with attributes, where the weight associated to edges corresponds to the ``distance''
among connected regions, while attributes specify their surface dimension. These new features are necessary in order to outline the spatial distribution of patches and the relevant additional
features associated to them: the dimension of a patch is needed to
define the density of the populations living inside that patch, while the distance is
needed to identify isolated patches, as well as to define the dispersal rates
of migrating individuals.
Moreover, by using some rules which do not modify the objects on which
they act (the so-called \virg mute rules''), we modified the classical view of maximal parallelism, by
allowing the maximal application of rules but, at the same time, reducing the maximal
consumption of objects. The model was applied to investigate some emergent metapopulation
behaviors, such as the influence of patch dimension, patch-to-patch distance, stochastic breeding, the dynamics underlying migration and colonization, the effects due to isolated patches, etc.
Then, in \cite{bicta07} we extended the analysis of that model by focusing on periodic resource feeding strategies, and  compared different systems where either increasing, decreasing, stationary or purely feeding stochastic phases were
defined inside each patch. We have shown there, for instance, how the seasonal
variance can transform the basic Lotka-Volterra dynamics inside each patch into
a more complex dynamics, where the different phases of a feeding cycle can be
identified through the effect that they have on the standard oscillations of preys and predators.

\begin{figure}[htbp]
\begin{center}
\includegraphics[width=10cm]{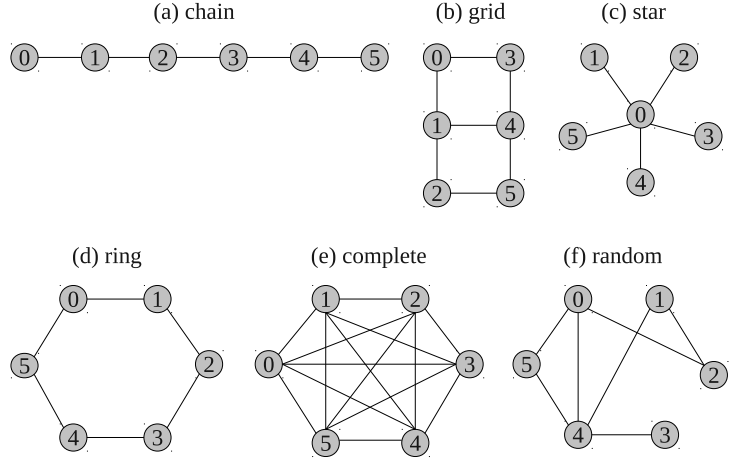}
\end{center}
\caption{Network topologies.} \label{fig:topologies}
\end{figure}

In this section, we present a simplified model of metapopulations, which exploits the multivolume stochastic simulation algorithm tau-DPP \cite{tauWMC7,BioSimWare}. With respect to the previous model, here we will not need to use the concept of mute rules, as the probabilistic choice and applications of rules is already embedded in the tau leaping algorithm \cite{Gill06}, on which tau-DPP is based. Moreover, we will not consider the presence of the dispersal pool, but we will instead focus our analysis on the direct communication of individuals among interconnected patches, according to some fixed network topologies. In order to compare the influence of each network, we have decided to perform our analysis on a total of 6 patches, spatially arranged in different ways. Namely, we assume that these network topologies can be described by graphs having the same number of nodes, but distinct connections, such as the chain, grid, star, ring, complete or random structure (see graphs $a, b, c, d, e, f$, respectively, in Fig. \ref{fig:topologies}). From now on, we will refer to the formal data structure by using the term `graph', and use the term `network' to denote the topological relationship on each graph.

%%%%%%%%%%%%%%%%%%%%%%%%%%%%%%%%%%%%%%%%%%%%%%%%%%%%%%%%%
%%%%%%%%%%%%%%%%%%%%%%%%%%%%%%%%%%%%%%%%%%%%%%%%%%%%%%%%%

Formally, each network topology $\nu \in \{a, b, c, d, e, f\}$, can be generally described by a weighted undirected graph $G_{\nu}=(N_{\Delta}^{\nu}, E^{\nu}, w^{\nu})$ where:
\begin{itemize}
    \item $N_{\Delta}^{\nu}$ is the set of nodes, such that each node $p_i \in N_{\Delta}^{\nu}$, $i$=1, $\dots$, 6, is characterized by a value $\delta(p_i) \in \Delta$ (with $\Delta$ being a set of attributes of some kind);
    \item $E^{\nu} \subseteq \{(p_i,p_j) \mid p_i,p_j \in N_{\Delta}^{\nu}\}$ is the set of (undirected) edges between nodes;
    \item $w^{\nu} : E^{\nu} \ra \mathbb{R}^+$ is the weight function associating a cost to
    each edge.
\end{itemize}

In the case of metapopulations, the set of nodes $N_{\Delta}^{\nu}$ coincides with the set of patches, the attribute of a node represents the area of the patch, the edges characterize which patches are directly
reachable from any patch (self-edges might exist as well but will not be considered in this work), and the weight $w^{\nu}_{i,j}$ of an edge $(p_i, p_j)$ represents a cost to measure the effort that individuals have to face when moving from patch $p_i$ to $p_j$. Given a network topology $\nu$, we denote by $Adj(p_i)^{\nu}$ the set of nodes that are directly connected to any node $p_i$, that is, $Adj(p_i)^{\nu}=\{p_j \in N_{\Delta}^{\nu} \mid \exists \: (p_i,p_j) \in E^{\nu}\}$. We also denote by $deg(p_i)^{\nu}$ the degree of patch $p_i$, that is, the number of patches directly connected to $p_i$ (formally, $deg(p_i)^{\nu}=card(Adj(p_i)^{\nu})$). We outline that, in what follows, we will assume that: (1) $w^{\nu}_{i,j}=1$ for each $(p_i,p_j) \in E^{\nu}$ and each $\nu \in \{a, b, c, d, e, f\}$, that is, all edges have the same cost; (2) $\delta(p_i)=1$ for each $p_i \in N_{\Delta}^{\nu}$ and each $\nu \in \{a, b, c, d, e, f\}$, that is, all patches have the same dimension. The rational behind this is that, in this paper, we focus our attention on the influence that different topologies of the habitat network can have on the local and global dynamics of metapopulations, regardless of the local features of each patch, or of the distances between patches. These features might be naturally added in further works related to this model, where real data can be used to define a specific model of some metapopulation systems.

In addition to the chosen network topology, this model of metapopulations also considers the presence of species individuals, which locally interact according to a chosen dynamics, and give rise to global dynamics thanks to the dispersal processes. To this purpose, in this paper we assume that each patch is characterized by the Lotka-Volterra (LV) model describing the interaction between the individuals of two populations, namely preys and predators. Inside each patch, the LV model is described by the following set of internal rules:
%$$r_1: AX \ra XX, r_2: XY \ra YY, r_3: Y \ra \lambda,$$
\begin{eqnarray*}
  r_1 :& AX \ra XX \\
  r_2 :& XY \ra YY \\
  r_3 :& Y \ra \lambda 
\end{eqnarray*}
\noindent where $X$ denotes the preys, $Y$ denotes the predators, $A$ denotes the sustenance resources and $\lambda$ is the empty symbol. Rules $r_1$ and $r_2$ model the growth of preys and predators, respectively, while rule $r_3$ models the death of predators. Each rule is also characterized by a stochastic constants (expressed in $time^{-1}$), that is used -- together with the current amounts of individuals occurring in the patch -- to evaluate its application probability step by step, according to the tau leaping algorithm (see \cite{Gill06,tauWMC7,VolumeRozenberg} for more details). All the simulations shown hereafter have been executed using the following values of stochastic constants and of initial amount of preys, predators, and sustenance resources: $c_1$=0.1, $c_2$=0.01, $c_3$=10, $X_0$=$Y_0$=1000, $A_0$=200 (the value of $A$ is fixed for the entire duration of each simulation). The simulations have been performed with the software BioSimWare \cite{BioSimWare}, that implements different stochastic simulation algorithms for both single and multivolume systems. The software is available for free download at http://bimib.disco.unimib.it/index.php/Software.

\begin{figure}[htbp]
\begin{center}
\includegraphics[width=8cm]{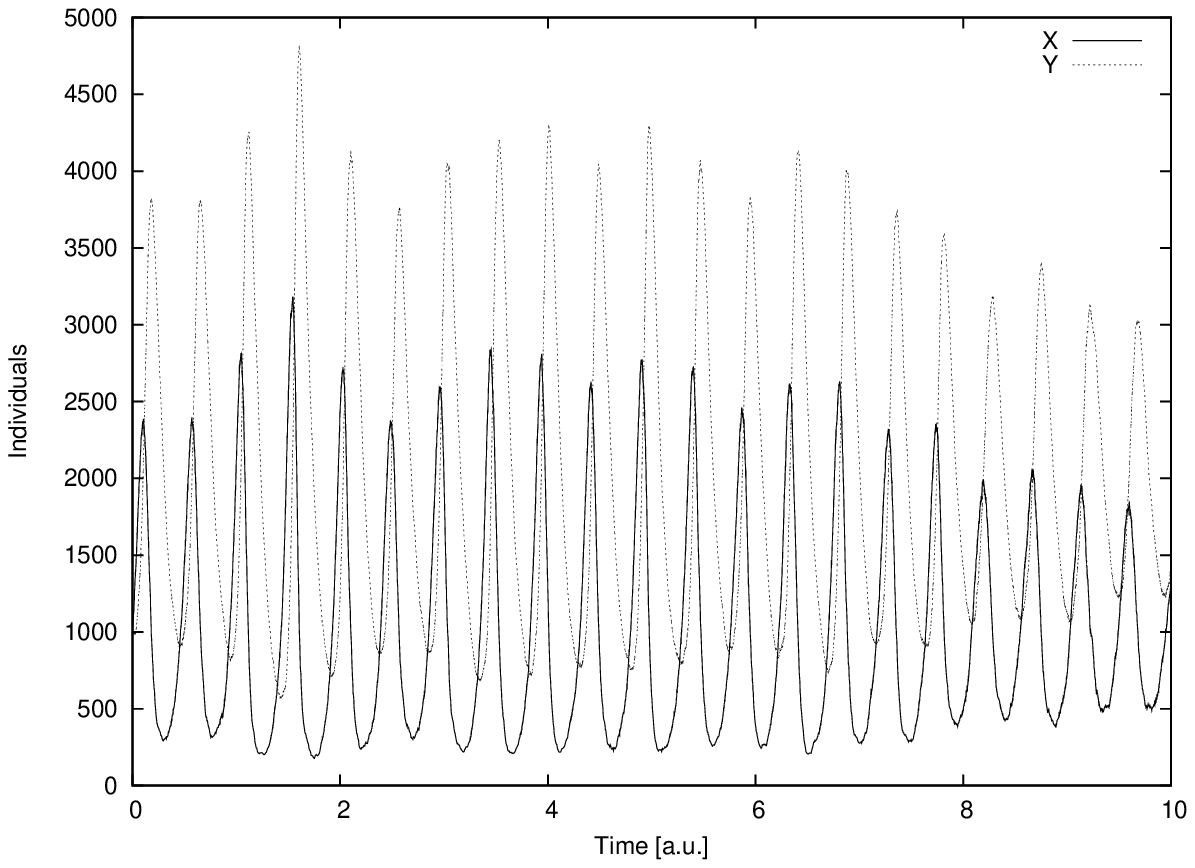} \hspace{-0.5cm}
\includegraphics[width=8cm]{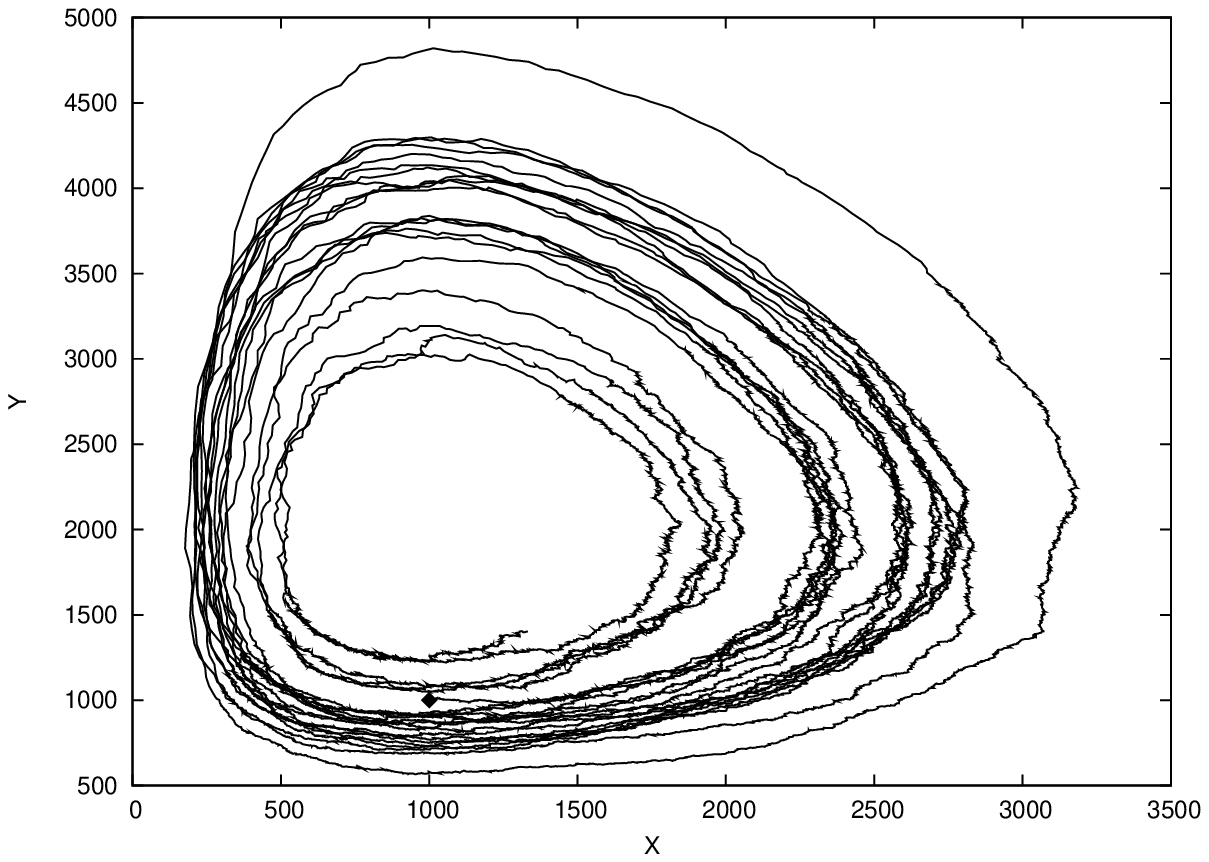}
\end{center}
\caption{The Lotka-Volterra dynamics in the single patch: oscillations in preys, $X$, and predators, $Y$ (left side), and corresponding phase space (right side).} \label{fig:LV_singlepatch}
\end{figure}

In Fig. \ref{fig:LV_singlepatch} we show the oscillating dynamics (left side) of preys and predators in the single patch, obtained with this choice of parameters, and the corresponding phase space (right side). These figures can be considered as reference to compare and discuss the dynamics obtained in the multi-patch model, as described in Section \ref{sec:sim_dyn}.

%%%%%%%%%%%%%%%%%%%%%%%%%%%%%%%%%%%%%%%%%%%%%%%%%%%%%%%%%
%%%%%%%%%%%%%%%%%%%%%%%%%%%%%%%%%%%%%%%%%%%%%%%%%%%%%%%%%

The single patch model is then extended to a multi-patch model where, inside each patch $p_i$ of each network topology $\nu$, we add as many communication rules as the number of patches connected to $p_i$ (that is, a total of $deg(p_i)^{\nu}$ rules inside each patch). These rules are needed to move population individuals among the various patches of the network, thus allowing to analyze the effects of migration and colonization in the metapopulation. This is done by attaching a destination target to each communication rule, specifying the destination patch, as it is usually done in P systems. Formally, in each patch $p_i$ of network $\nu$, we add the so-called \emph{dispersal rules}
$$r_{d_{p_j}} : Y \ra (Y, target(p_j)),$$
for each $p_j \in Adj(p_i)^{\nu}$. Similarly to the local rules $r_1, r_2, r_3$, the probability of applying each dispersal rule is determined by using its stochastic constant $c_{d_{p_j}}$, whose values will be given in the next section to consider different migration rates.

\section{The influence of network topologies on metapopulation dynamics}\label{sec:sim_dyn}

In this section we analyze how the topological structure of patch-to-patch connections, and the rate of individual dispersal between connected patches, influence the local and global dynamics of a metapopulation. In particular, in Section \ref{subsec:communication} we consider how a given topology and a fixed dispersal rate can influence the prey-predators dynamics, while in Section \ref{subsec:colonization} we focus on the capability of colonization of empty patches, starting from the dispersal of predators living in a few patches which occupy peculiar positions in the given network topology.

\subsection{Network topologies and migration}\label{subsec:communication}

In this section, we analyze the role of migration and compare the six network topologies with respect to four different conditions for the dispersal rules. Namely, we assume that each patch of each topology is initialized with a complete LV model as given in Section \ref{subsec:Psyst_metapop}, where the value of the stochastic constant $c_{d_{p_j}}$ for the dispersal of predators, in each patch $p_i \in N_{\Delta}^{\nu}$, can assume one of the following values:
\begin{enumerate}
\item $c_{d_{p_j}}$=1, for each $p_j \in Adj(p_i)^{\nu}$;
\item $c_{d_{p_j}}$=10, for each $p_j \in Adj(p_i)^{\nu}$;
\item $c_{d_{p_j}}$=20, for each $p_j \in Adj(p_i)^{\nu}$;
\item $c_{d_{p_j}}$=$\frac{10}{deg(p_i)}$, for each $p_j \in Adj(p_i)^{\nu}$.
\end{enumerate}
\smallskip
By considering the first condition as reference, the power of dispersal in the second (third) condition is ten-fold (twenty-fold) the first one, irrespective of the position that patch $p_i$ occupies in the considered network. In other terms, the flux of dispersal from each patch, in the first three conditions, results amplified by the number of connections that each patch has with respect to the other patches in the network. On the contrary, the fourth condition corresponds to the situation when, for each patch $p_j \in Adj(p_i)^{\nu}$, the sum of the values of constants of dispersal rules in $p_i$ is always equal to 10, but the rate of dispersal along each edge from $p_i$ to $p_j$ depends on the degree of $p_i$. For instance, in the network topology $a$ (Fig. \ref{fig:topologies}), the value of $c_{d_{p_j}}$ in patches $p_0$ and $p_5$ is equal to 10, while the value of $c_{d_{p_j}}$ in patches $p_1$, $\dots$, $p_4$ is equal to 5; in the network topology $c$ (Fig. \ref{fig:topologies}), the value of $c_{d_{p_j}}$ in patch $p_0$ is equal to 2, while the value of $c_{d_{p_j}}$ in all other patches is equal to 10, and so on. So doing, we can weigh the dispersal of predators according to the position of each patch in the network, and simulate a situation where the flux of dispersal from each patch towards its adjacent patches is uniform throughout the whole network.

%%%%%%%%%%%%%%%%%%%%%%%%%%%%%%%%%%%%%%%%%%%%%%%%%%%%%%%%%
\begin{figure}[htbp]
\begin{center}
\includegraphics[width=8cm]{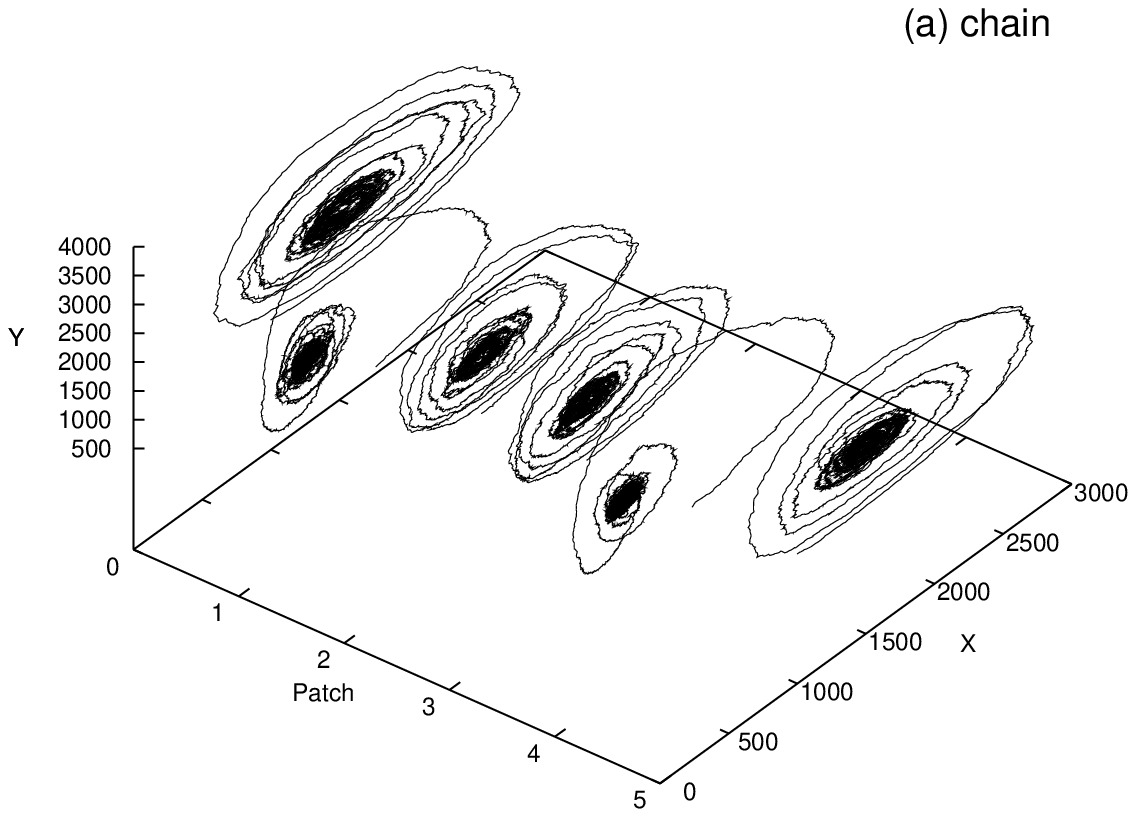} \hspace{-0.8cm}
\includegraphics[width=8cm]{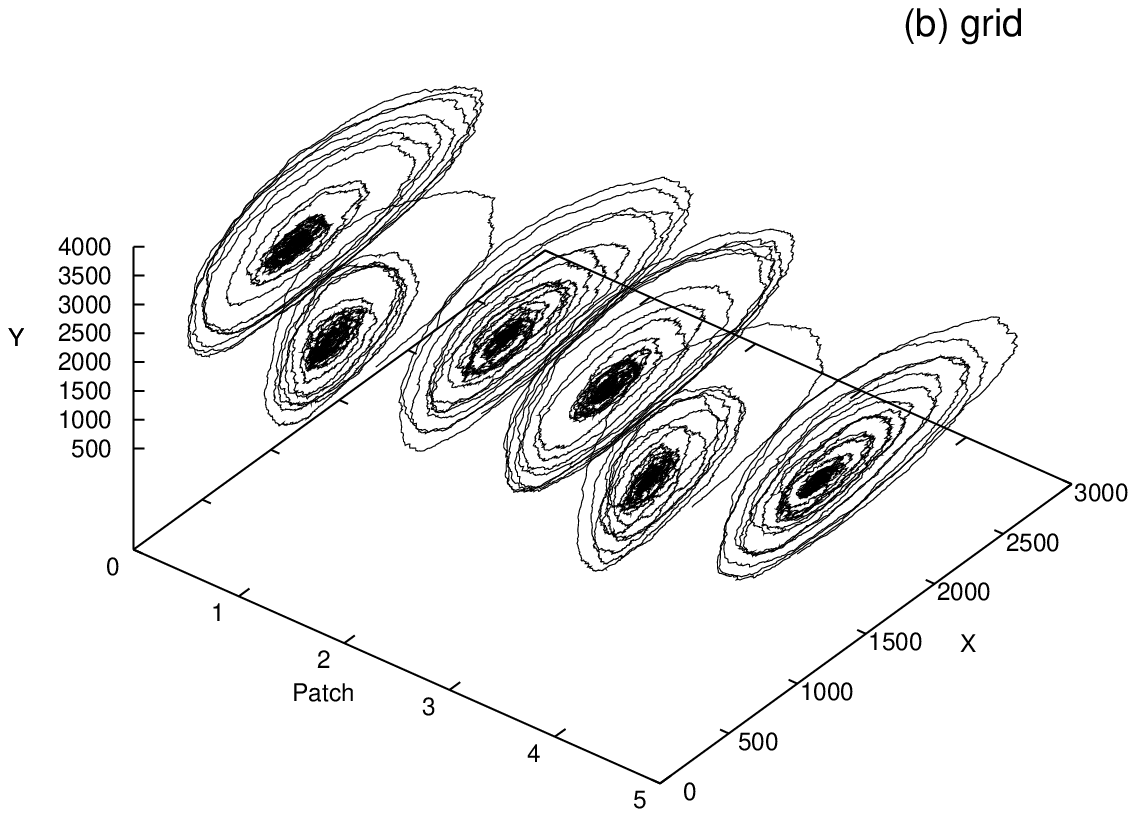}
\includegraphics[width=8cm]{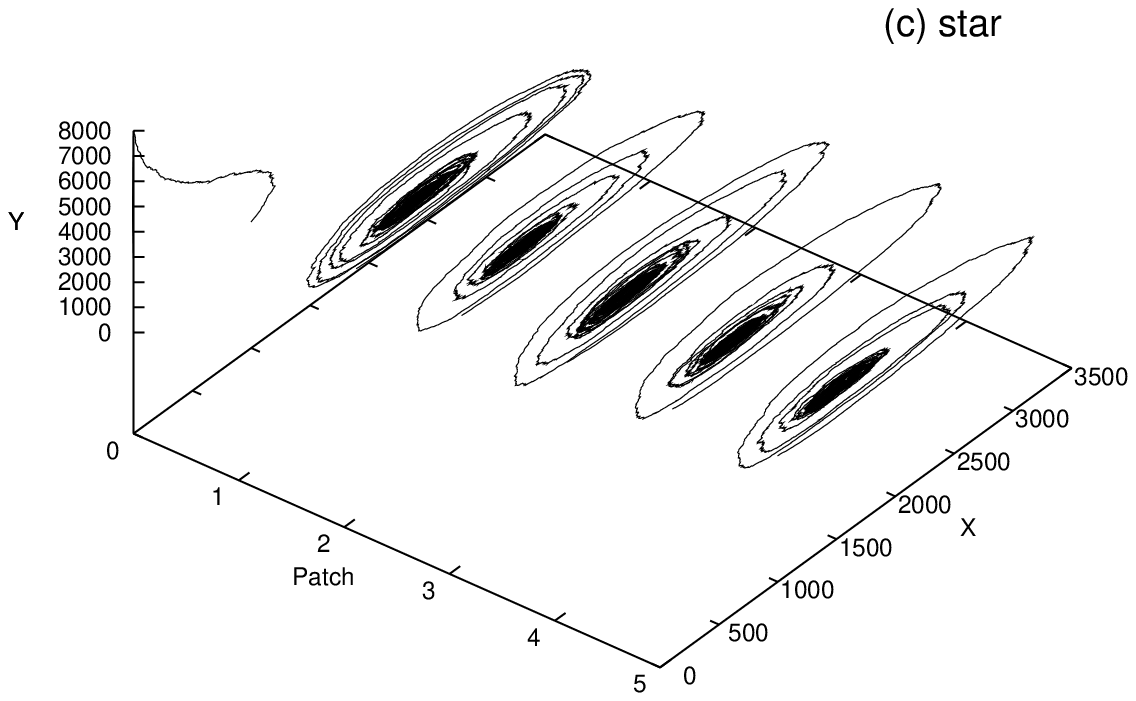} \hspace{-0.8cm}
\includegraphics[width=8cm]{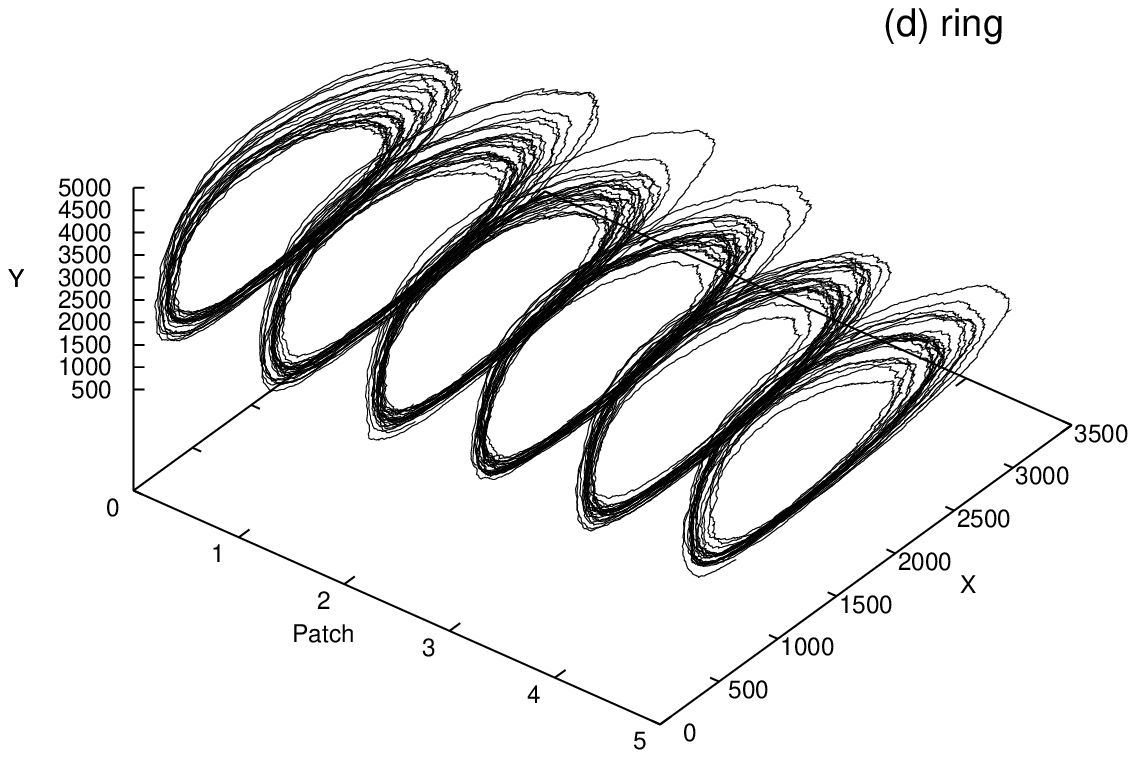}
\includegraphics[width=8cm]{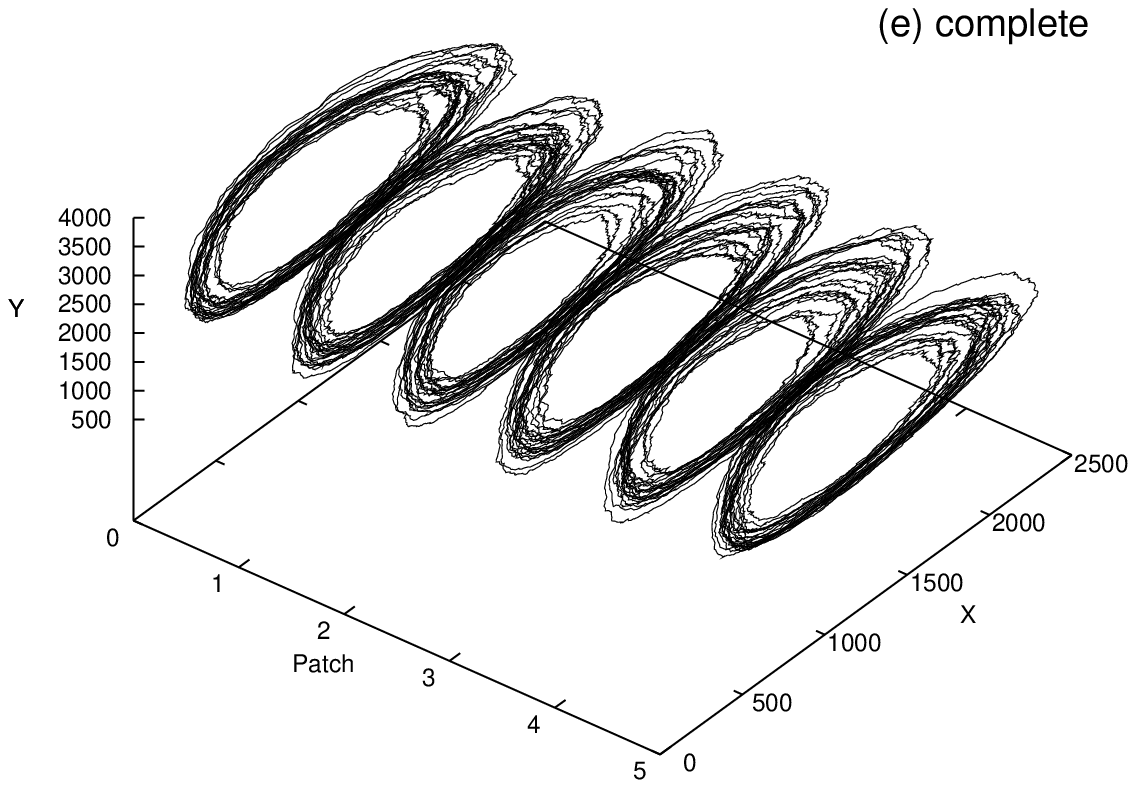} \hspace{-0.8cm}
\includegraphics[width=8cm]{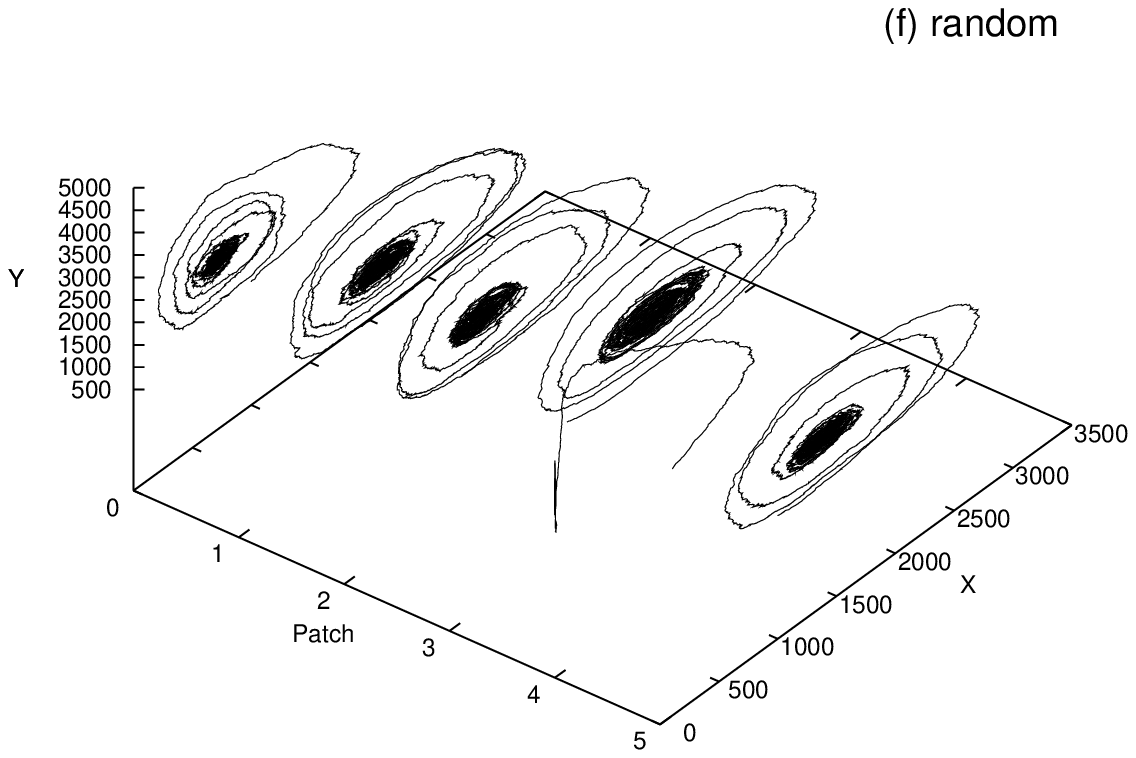}
\end{center}
\caption{The power of migration: LV dynamics in the phase space of each network topology.} \label{fig:phasespace_networks}
\end{figure}
%%%%%%%%%%%%%%%%%%%%%%%%%%%%%%%%%%%%%%%%%%%%%%%%%%%%%%%%%

For space limits, in Fig. \ref{fig:phasespace_networks} we present the phase spaces of all network topologies, obtained from simulations of the fourth condition only. For each network, in particular, we show the phase space of the local dynamics of each patch. The graphics show that, in the case of the chain graph (phase space (a)), the patches having different degrees are characterized by different dynamics: in fact, patches $p_0$ and $p_5$ show a different behavior with respect to the other patches. In addition to the role of patch degree, we can see that also the position of patches in the graph plays a central role: despite the fact that patches $p_1, p_2, p_3$ and $p_4$ have all the same degree, the dynamics inside $p_1$ and $p_4$ differs from that of patches $p_2$ and $p_3$. This is due to the different power of dispersal rules of their two neighbors, namely $c_{d_{p_j}}=10$ in patches $p_0$, $p_5$, while $c_{d_{p_j}}=5$ in patches $p_2$, $p_3$, which cause a larger flux of predators dispersal towards patches $p_1$ and $p_4$. The global effect is the presence of three different dynamics (one in $p_0$, $p_5$, another one in $p_1$, $p_4$, and a third one in $p_2$, $p_3$), all of which are characterized by oscillations in $X$ and $Y$ with no regular amplitudes (compare these phase spaces with the standard LV phase space in the single patch model given in Fig. \ref{fig:LV_singlepatch}, right side, and also with the phase spaces in Fig. \ref{fig:phasespace_networks}, graphics (d) and (e)). Furthermore, we can evidence that these oscillations are characterized by an initial wider amplitude, which is reduced during time.

Similarly, the dynamics of the patches in the grid graph (phase space (b)) is influenced only by the number of edges; in this phase space, we can identify two different types of dynamics: one for the patches with three edges ($p_1$, $p_4$) and another one for those with two connections.

In the star graph (phase space (c)), the LV dynamics endures in all patches apart from $p_0$, where the number of preys $X$ collapses to an attractor in zero, and no oscillations according to the LV dynamics in both $X$ and $Y$ can be established. In this patch, the number of predators fluctuates in a certain range, because of their dispersal from/to the other patches. Basically, in this condition patch $p_0$, that represents the center of the star, becomes a local area of the habitat where only dispersal occurs.

The simulations for the ring and complete graphs (phase spaces (d), (e)) show very similar results: in both cases, all patches in each graph have the same degree (two in the first configuration and five in the second one), leading to regular oscillations in $X$ and $Y$ with almost constant amplitude.

The results concerning the last configuration, the random graph (phase space (f)), show a combination of the effects described above. In particular, the dynamics of the patches differ each other depending on the degree of the patches themselves; moreover, in $p_4$, which is characterized by the highest degree, the high number of incoming predators (migrating from the four adjacent patches) leads to the extinction of preys (similarly to what happens in patch $p_0$ of the star graph).

We also tested, for each network topology, the other three conditions listed above. In these cases, the results have shown that the amplification of the power of dispersal with respect to the patch degree gives rise to a balance between the incoming and migrating individuals, which leads to comparable LV dynamics for all networks, with regular oscillations inside each patch (data not shown).

\subsection{Network topologies and colonization}\label{subsec:colonization}

In this section, we compare the six network topologies with respect to the capability of colonizing the empty patches that each network contains, starting from the patches that contain a complete LV model and that occupy a peculiar position in that network . We recall that in this work we are considering only the migration of predators, hence the empty patches are hereby assumed to contain no predators but only an initial amount of preys. In each network $\nu$, the set of patches initialized with the complete LV model will be denoted as $p_{LV}^{\nu}$. To test the feature of colonization, we consider four different initial conditions, hereby denoted as IC$k$, $k$=$1, \dots, 4$, where $Y_0$=0 and:
\begin{enumerate}
\item IC1 is characterized by $c_{d_{p_j}}$=1 and $X_0$=10;
\item IC2 is characterized by $c_{d_{p_j}}$=1 and $X_0$=100;
\item IC3 is characterized by $c_{d_{p_j}}$=10 and $X_0$=10;
\item IC4 is characterized by $c_{d_{p_j}}$=10 and $X_0$=100.
\end{enumerate}
In each given network, all empty patches are initialized with the same chosen condition IC$k$, besides the patches in the set $p_{LV}^{\nu}$ that are initialized with a standard LV model, having the communication constant $c_{d_{p_j}}$ equal to the one given in the chosen IC$k$, and all other parameters as given in Section \ref{subsec:Psyst_metapop}.

With this type of analysis, we expect to determine which features of the network topologies are more relevant with respect to the colonization of empty patches, under a given initial condition. All conditions have been tested for each network and, for each fixed initial condition, different sets of $p_{LV}^{\nu}$ have been considered. In the following, for space limits, we present only some results of these simulations, and briefly discuss the results obtained in the other analyzed conditions. In each of the following graph, preys ($X$) are represented with solid lines, while predators ($Y$) are represented with dashed lines.

\begin{figure}[htbp]
\begin{center}
\includegraphics[width=8cm]{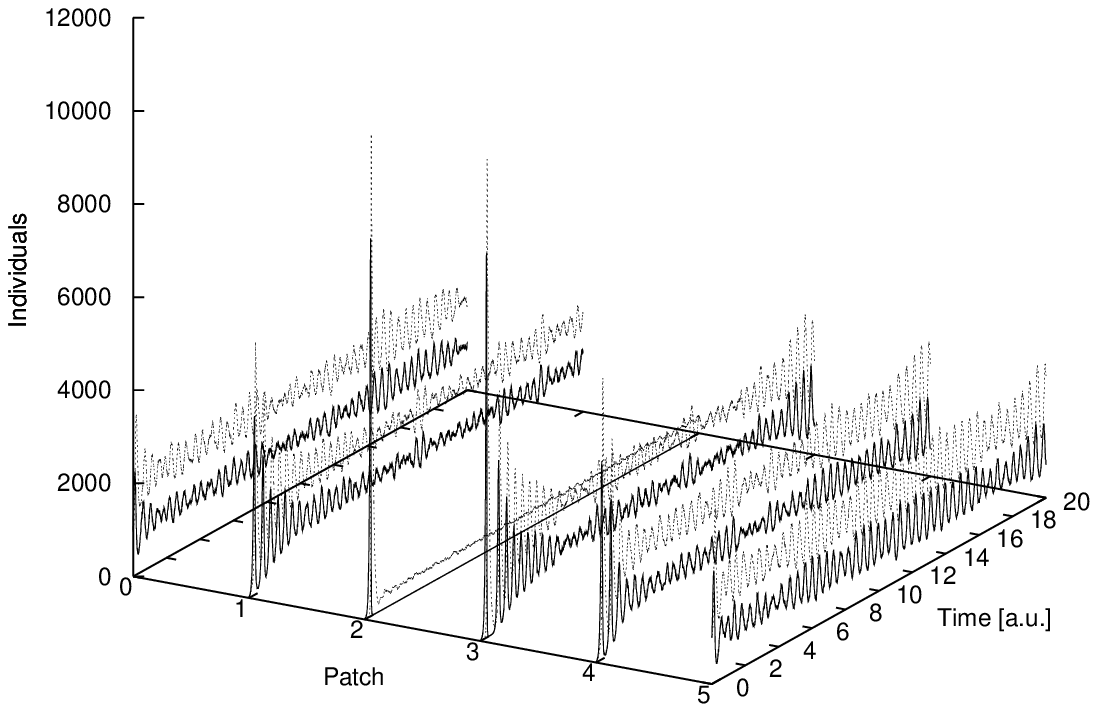} \hspace{-0.7cm}
\includegraphics[width=8cm]{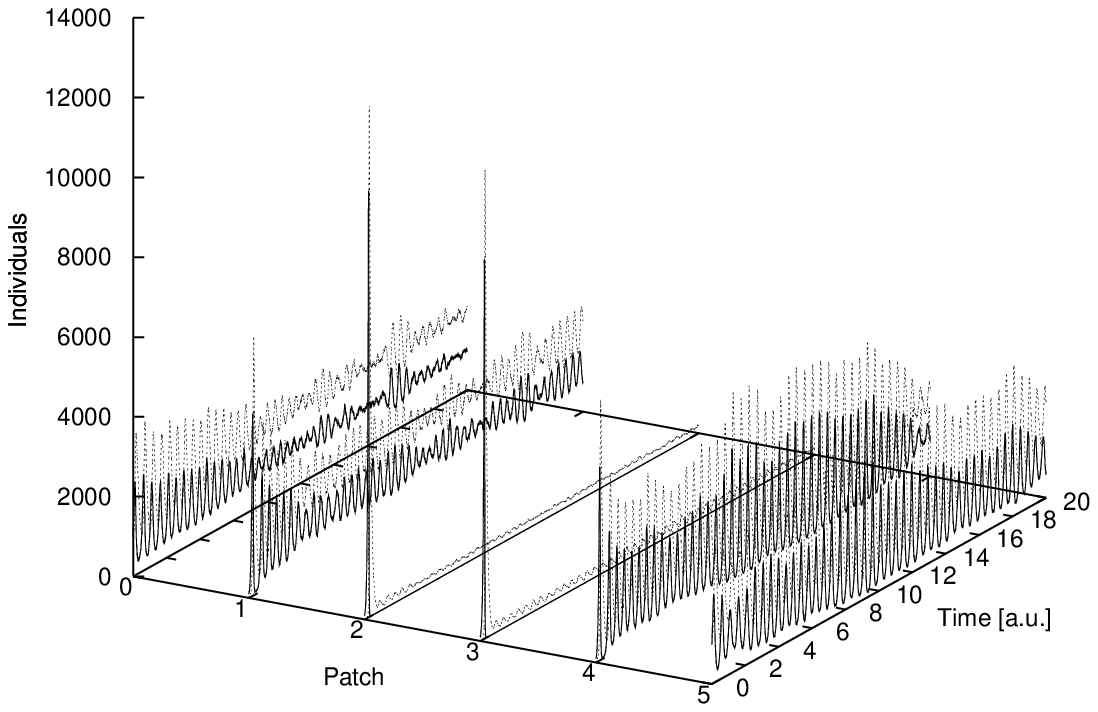}
\includegraphics[width=8cm]{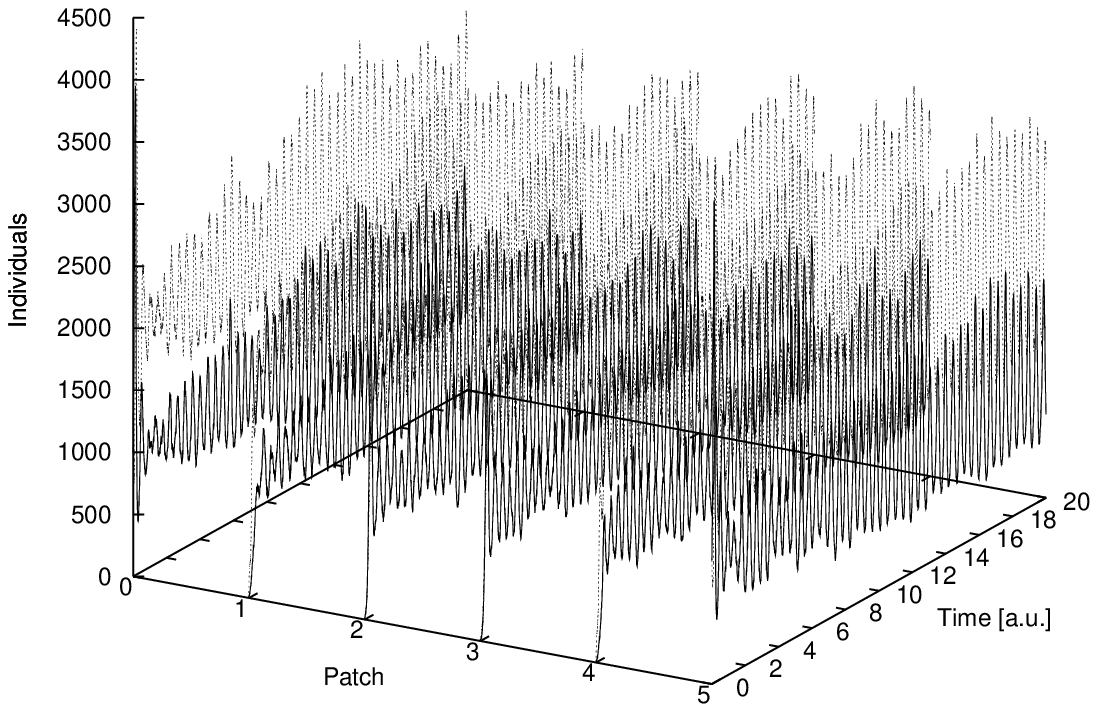} \hspace{-0.7cm}
\includegraphics[width=8cm]{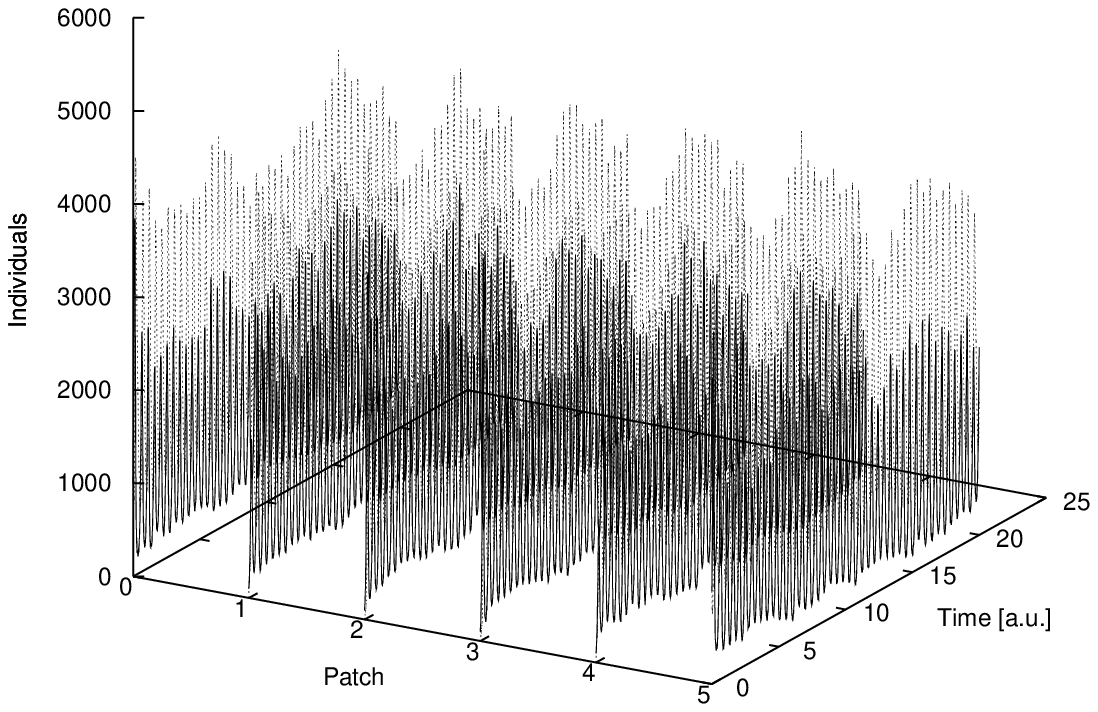}
\end{center}
\caption{Colonization in the chain topology, with $p_{LV}^{a}$=$\{p_0, p_5\}$ and initial conditions IC1 (top left), IC2 (top right), IC3 (bottom left), IC4 (bottom right).} \label{fig:chain_patches0and5}
\end{figure}

\begin{figure}[ht]
\begin{center}
\includegraphics[width=8cm]{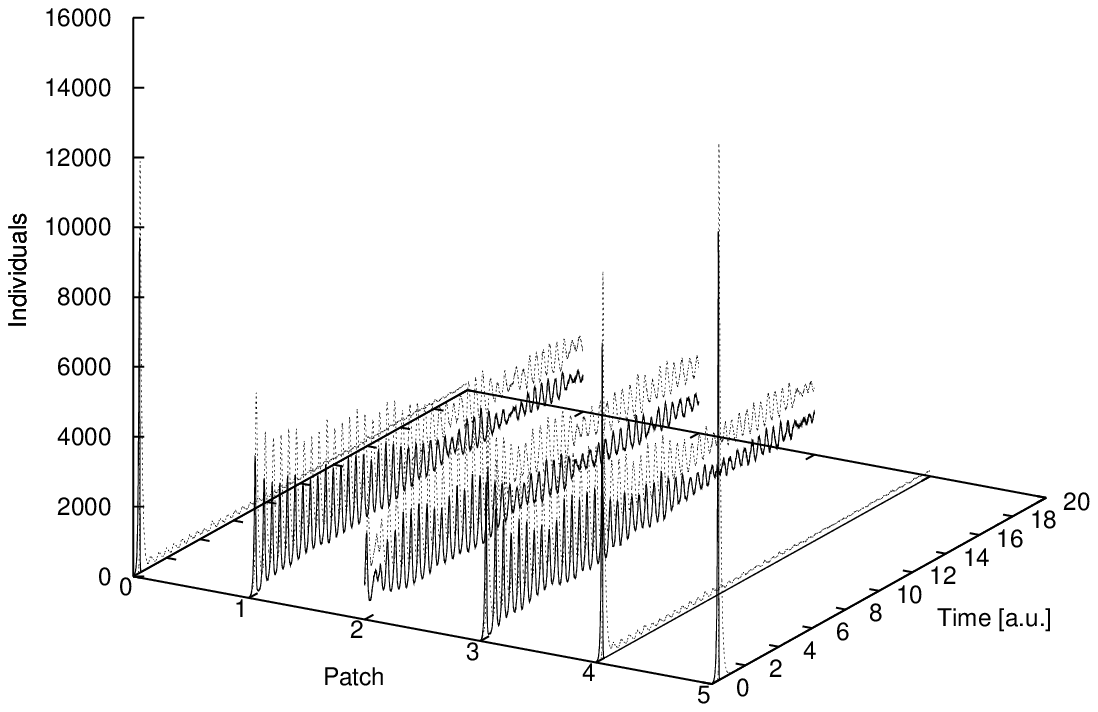} \hspace{-0.7cm}
\includegraphics[width=8cm]{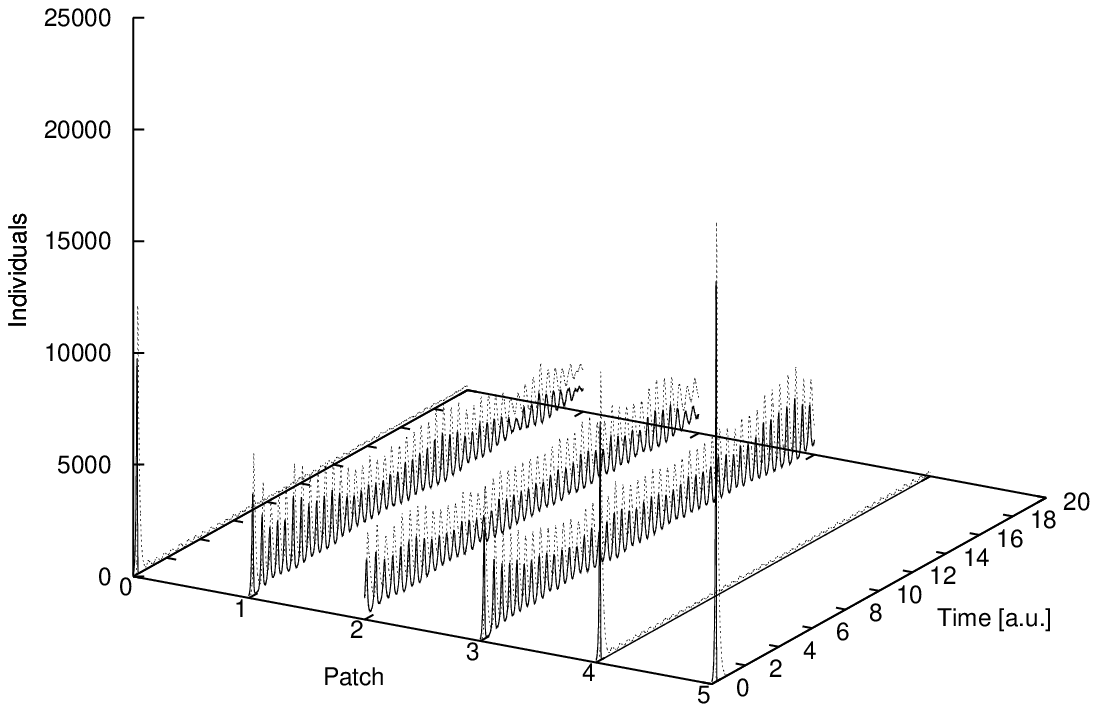}
\includegraphics[width=8cm]{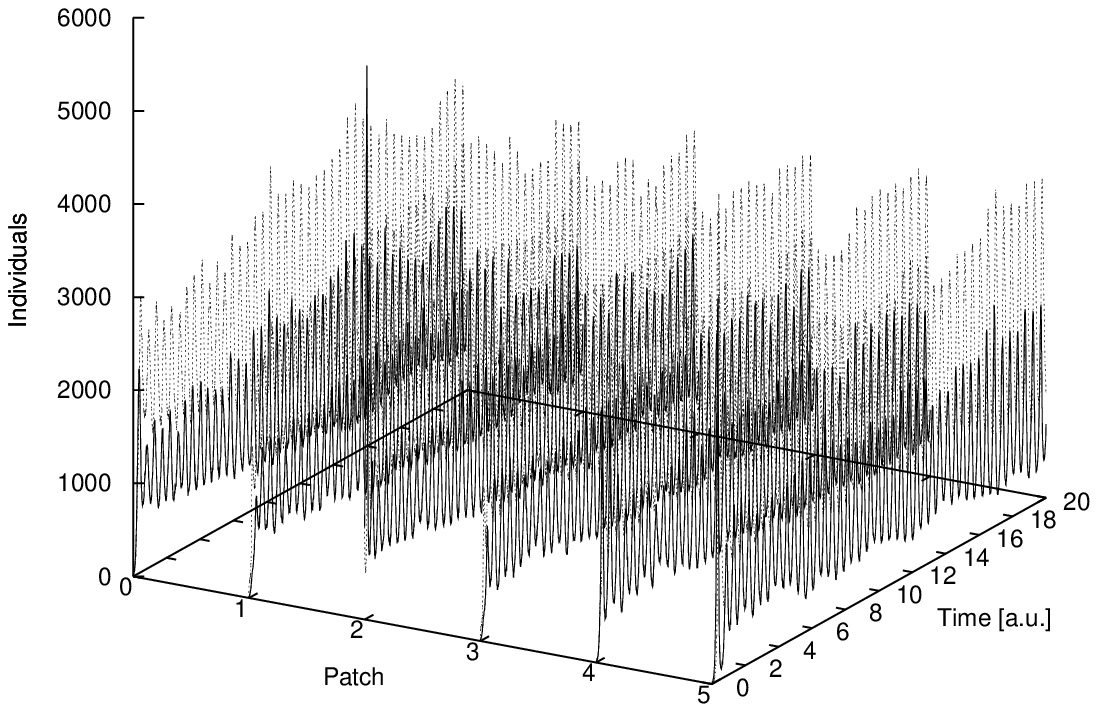} \hspace{-0.7cm}
\includegraphics[width=8cm]{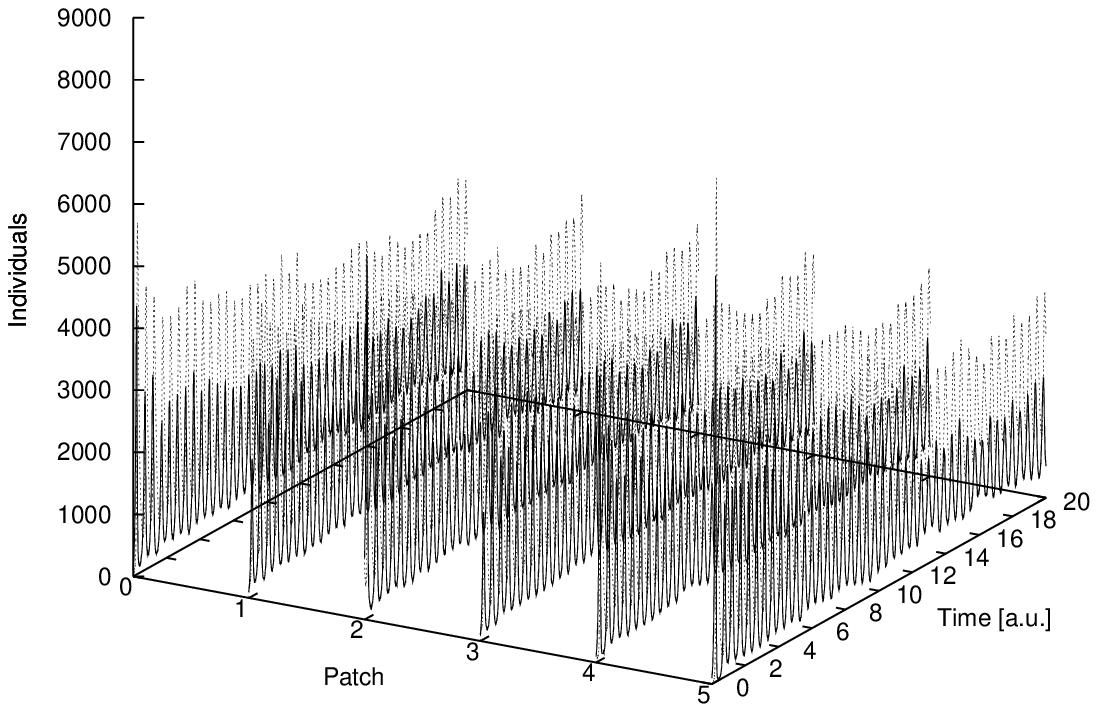}
\end{center}
\caption{Colonization in the chain topology, with $p_{LV}^{a}$=$\{p_2\}$ and initial conditions IC1 (top left), IC2 (top right), IC3 (bottom left), IC4 (bottom right).} \label{fig:chain_patch2}
\end{figure}

We start by considering the network $\nu=a$, that is, the chain graph. In this case, we present the results obtained in all the initial conditions IC$1$, IC$2$, IC$3$, IC$4$, considering three sets of LV patches, namely $p_{LV}^{a}$=$\{p_0, p_5\}$, $p_{LV}^{a}$=$\{p_2\}$  and  $p_{LV}^{a}$=$\{p_0\}$. In the first case ($p_{LV}^{a}$=$\{p_0, p_5\}$, shown in Fig. \ref{fig:chain_patches0and5}) we can see that, when the power of dispersal is low (IC1, IC2), the time required by the predators to reach the patches  $p_2$ and $p_3$, which are at the highest distance from $p_0$ and $p_5$, allows an initial uncontrolled growth of the preys in $p_2$ and $p_3$, which subsequently undergo extinction as soon as the predators enter the patch. Such \virg delay'' in the local establishment of a population of predators is the effect that prevent the formation of the LV dynamics; this effect, as shown hereafter, is a common aspect of all network topologies. Concerning the chain network, this is more evident in condition IC2, where the initial amount of preys inside the empty patches is higher than IC1: in this case, the LV dynamics can be established only in four of the six patches.
On the other hand, with the initial conditions IC3 and IC4, the power of dispersal is sufficient to colonize all of the patches, irrespectively of the numbers of preys that are initially present in the empty patches and of the position of the LV complete patch.
Similar results for the chain network have been obtained in the second analyzed case ($p_{LV}^{a}$=$\{p_2\}$, shown in Fig. \ref{fig:chain_patch2}) and in the third case ($p_{LV}^{a}$=$\{p_0\}$, data not shown).

%%%%%%%%%%%%%%%%%%%%%%%%%%%%%%%%%%%%%%%%%%%%%%%%%%%%%%%%%
%%%%%%%%%%%%%%%%%%%%%%%%%%%%%%%%%%%%%%%%%%%%%%%%%%%%%%%%%

%%%%%%%%%%%%%%%%%%%%%%%%%%%%%%%%%%%%%%%%%%%%%%%%%%%%%%%%%
%%%%%%%%%%%%%%%%%%%%%%%%%%%%%%%%%%%%%%%%%%%%%%%%%%%%%%%%%

\begin{figure}[htbp]
\begin{center}
\includegraphics[width=7.5cm]{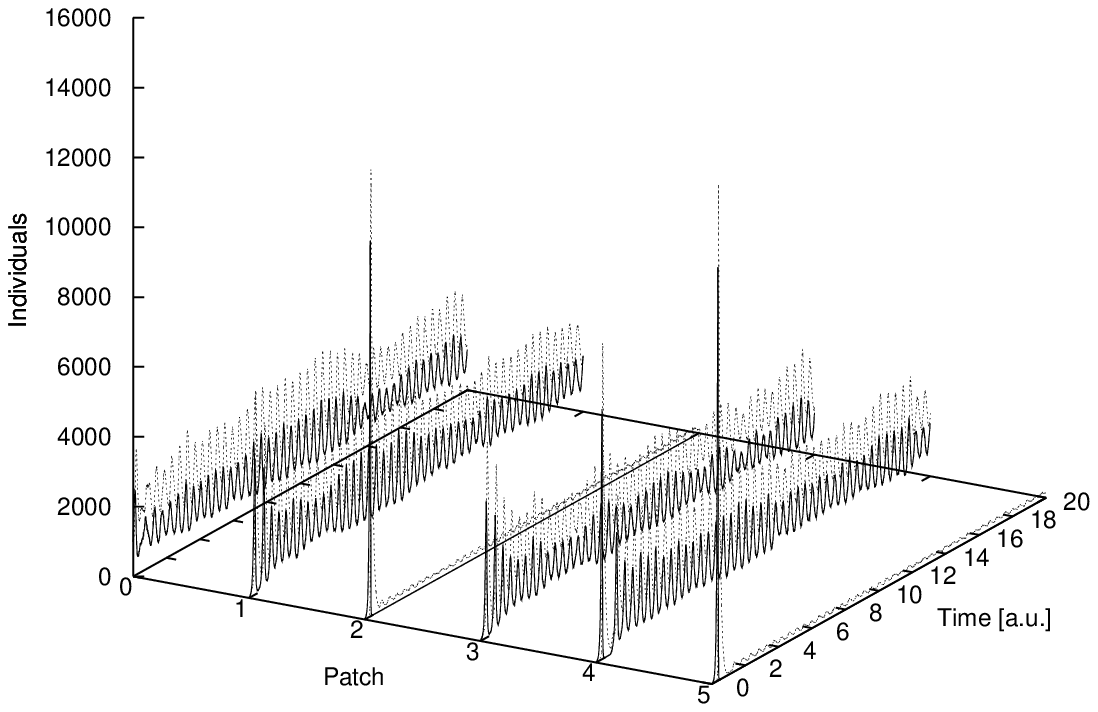} \hspace{-0.7cm}
\includegraphics[width=7.5cm]{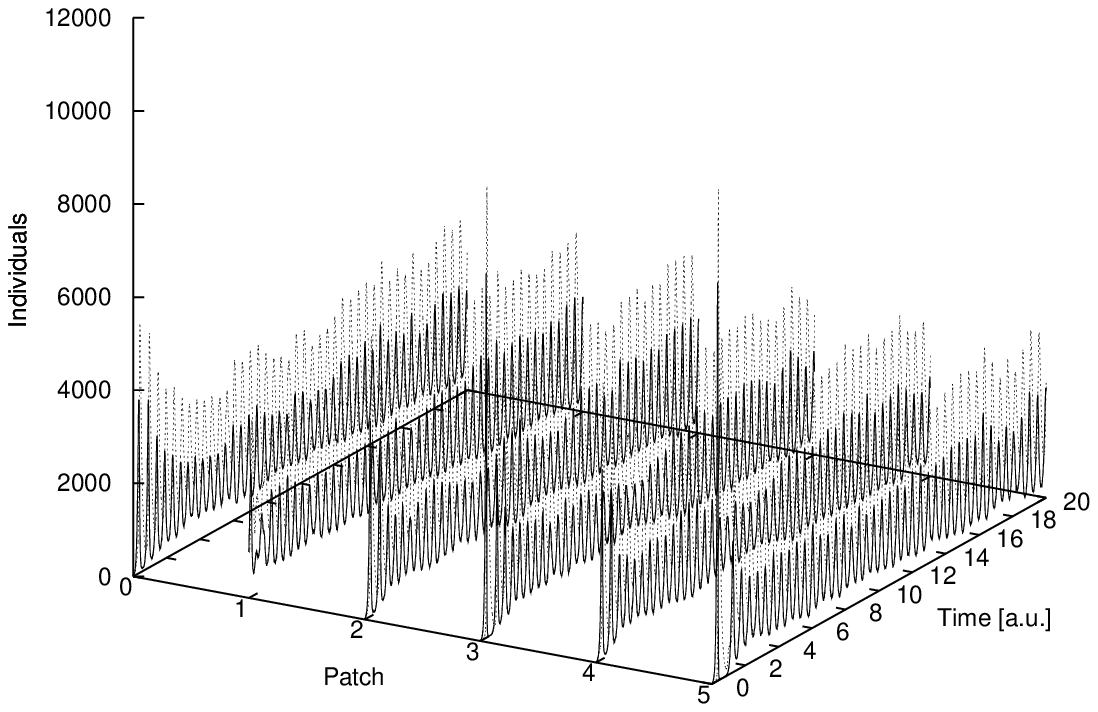}
\end{center}
\caption{Colonization in the grid topology, with initial condition IC1 and $p_{LV}^{b}$=$\{p_0\}$ (left), $p_{LV}^{b}$=$\{p_1\}$ (right).} \label{fig:grid}
\end{figure}

For the network topology $\nu=b$, that is, the grid graph, we show the results obtained in the cases IC$1$, when $p_{LV}^{b}$=$\{p_0\}$ (Fig. \ref{fig:grid}, left side) and $p_{LV}^{b}$=$\{p_1\}$ (Fig. \ref{fig:grid}, right side). According to the position of the LV complete patches in this network topology, we can see that, in the first case, the predators are capable to colonize patches $p_1$ and $p_3$, that are directly connected to $p_0$, and patch $p_4$, that is directly connected to both $p_1$ and $p_3$. However, patches $p_2$ and $p_5$ cannot be colonized. In the second case, the higher degree of the LV complete patch $p_1$, allows the colonization of all patches. With the initial condition IC2 (data not shown), in the other tested cases $p_{LV}^{b}$=$\{p_0\}$ and $p_{LV}^{b}$=$\{p_1\}$, only the patches directly connected to $p_0$ and $p_1$, respectively, are colonized by the predators.

%%%%%%%%%%%%%%%%%%%%%%%%%%%%%%%%%%%%%%%%%%%%%%%%%%%%%%%%%
%%%%%%%%%%%%%%%%%%%%%%%%%%%%%%%%%%%%%%%%%%%%%%%%%%%%%%%%%

%%%%%%%%%%%%%%%%%%%%%%%%%%%%%%%%%%%%%%%%%%%%%%%%%%%%%%%%%
\begin{figure}[htbp]
\begin{center}
\includegraphics[width=7.5cm]{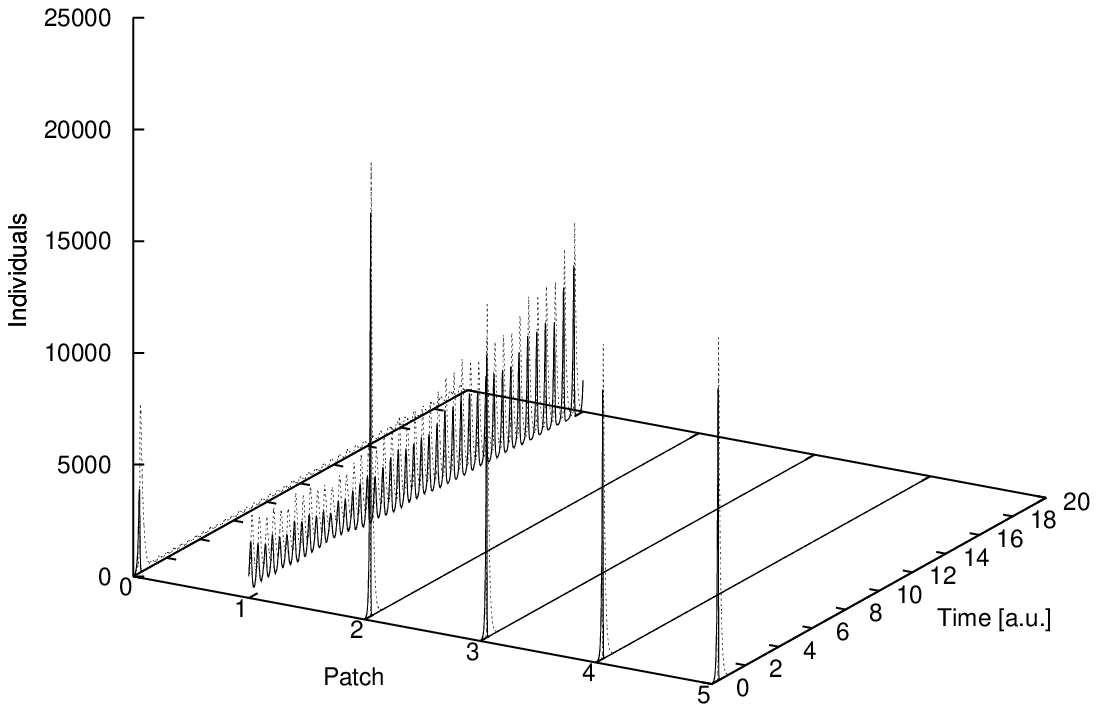} \hspace{-0.7cm}
\includegraphics[width=7.5cm]{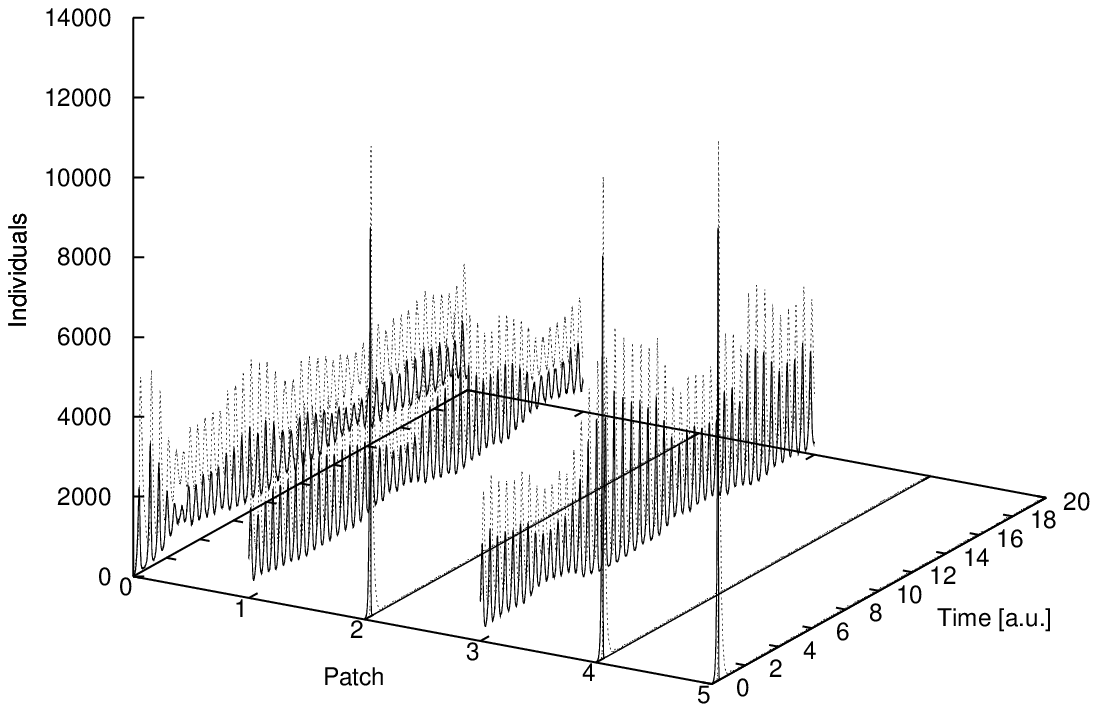}
\end{center}
\caption{Colonization in the star topology, with initial condition IC1 and $p_{LV}^{c}$=$\{p_1\}$ (left), $p_{LV}^{c}$=$\{p_1, p_3\}$ (right).} \label{fig:star}
\end{figure}
%%%%%%%%%%%%%%%%%%%%%%%%%%%%%%%%%%%%%%%%%%%%%%%%%%%%%%%%%

%%%%%%%%%%%%%%%%%%%%%%%%%%%%%%%%%%%%%%%%%%%%%%%%%%%%%%%%%
\begin{figure}[htbp]
\begin{center}
\includegraphics[width=7.5cm]{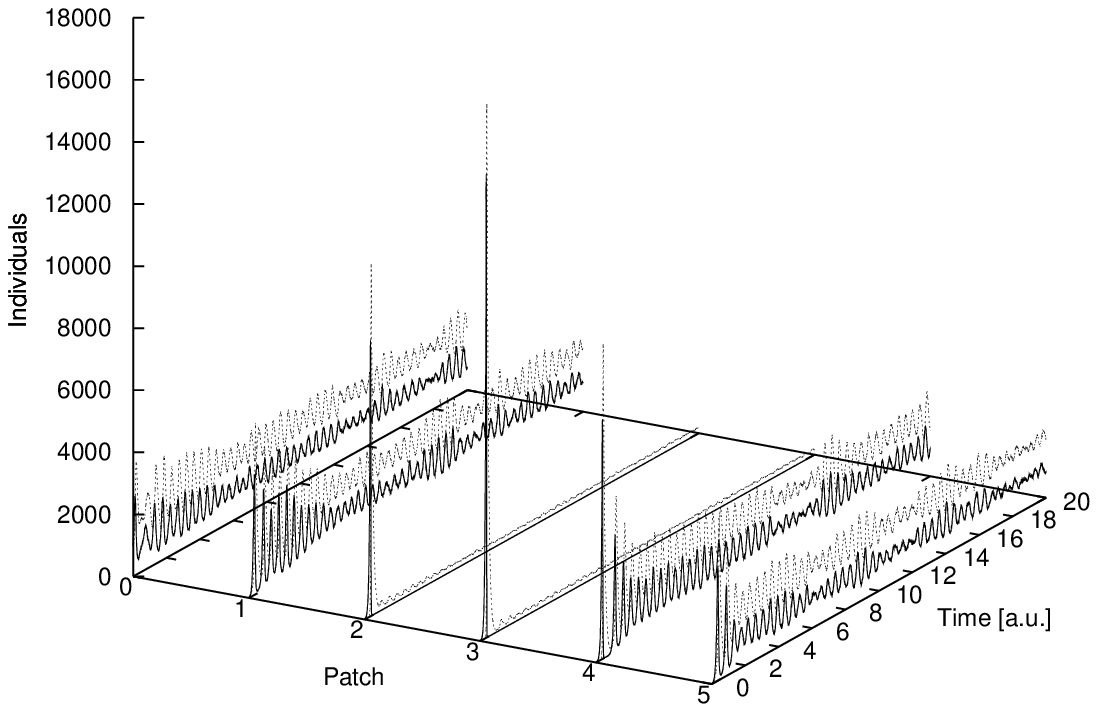} \hspace{-0.7cm}
\includegraphics[width=7.5cm]{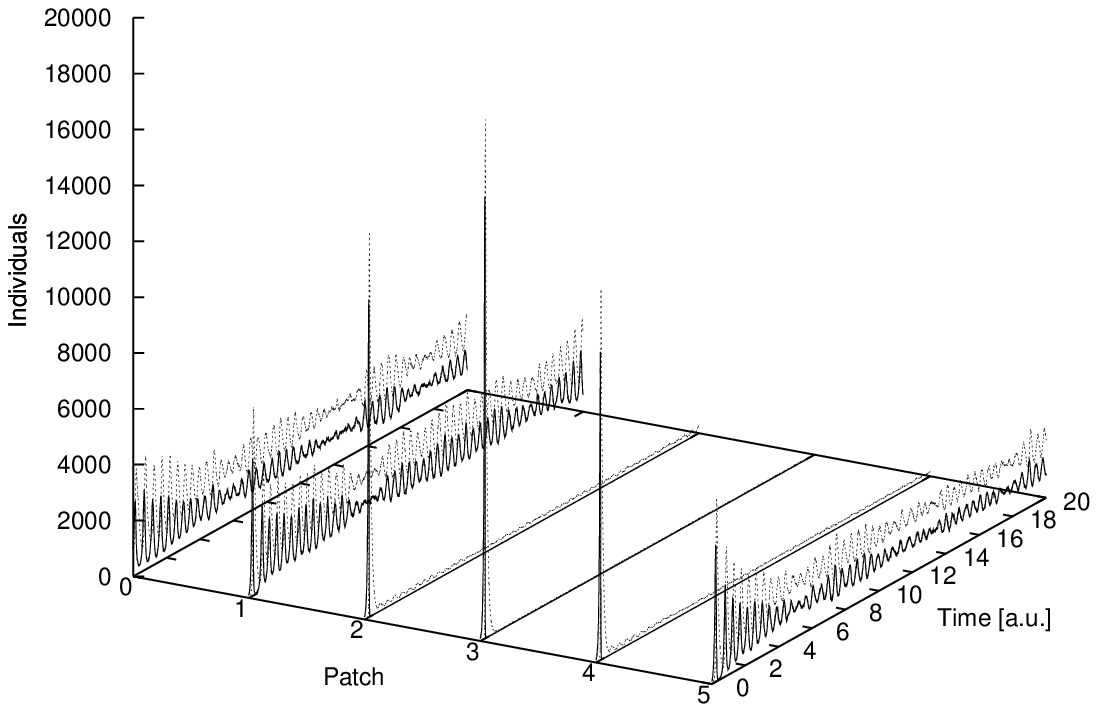}
\end{center}
\caption{Colonization in the ring topology, with $p_{LV}^{d}$=$\{p_0\}$ and initial condition IC1 (left) and IC2 (right).} \label{fig:ring}
\end{figure}
%%%%%%%%%%%%%%%%%%%%%%%%%%%%%%%%%%%%%%%%%%%%%%%%%%%%%%%%%

For the network topology $\nu=c$, that is, the star graph, we show the results obtained in the cases IC$1$, when $p_{LV}^{c}$=$\{p_1\}$ (Fig. \ref{fig:star}, left side) and $p_{LV}^{c}$=$\{p_1, p_3\}$ (Fig. \ref{fig:star}, right side). According to the position of the LV complete patches in this network topology, we can see that, in the first case, no patches are colonized because of the high degree of $p_0$ (which is the only patch connected to $p_1$) that spreads the predators over the other patches, thus preventing the formation of the LV dynamics. In the second case, the combined effect of migration from $p_1$ and $p_3$ allows the colonization of patch $p_0$, which is directly connected with both of them. We then performed other simulations starting with conditions IC3 and IC4: in these cases, the higher value of $c_{d_{p_j}}$ allows the colonization of every patch (except from patch $p_0$) independently from the initial position of the LV complete patch (data not shown). On the contrary, when we assume $p_{LV}^{c}$=$\{p_0\}$, that is, the center of the star, then all patches are fully colonized, independently from the considered initial condition.

For the network topology $\nu=d$, that is, the circular graph, we show the results obtained in the cases IC$1$ and IC$2$, when $p_{LV}^{d}$=$\{p_0\}$ (Fig. \ref{fig:ring}, left and right sides, respectively). Starting with the initial condition IC2, the predators are capable of colonizing only the patches directly connected to the LV complete patch $p_0$, while in the case IC1, also patch $p_4$ (being at distance 2 from the LV complete patch) is colonized. These results highlight, in particular, another aspect that was more marginal in the other simulations: the stochastic nature of the communication process and of the growth of preys, which leads to the extinction of preys in patch $p_2$, while in patch $p_4$ it drives the local behavior to an oscillatory dynamics.

\begin{figure}[htbp]
\begin{center}
\includegraphics[width=7.5cm]{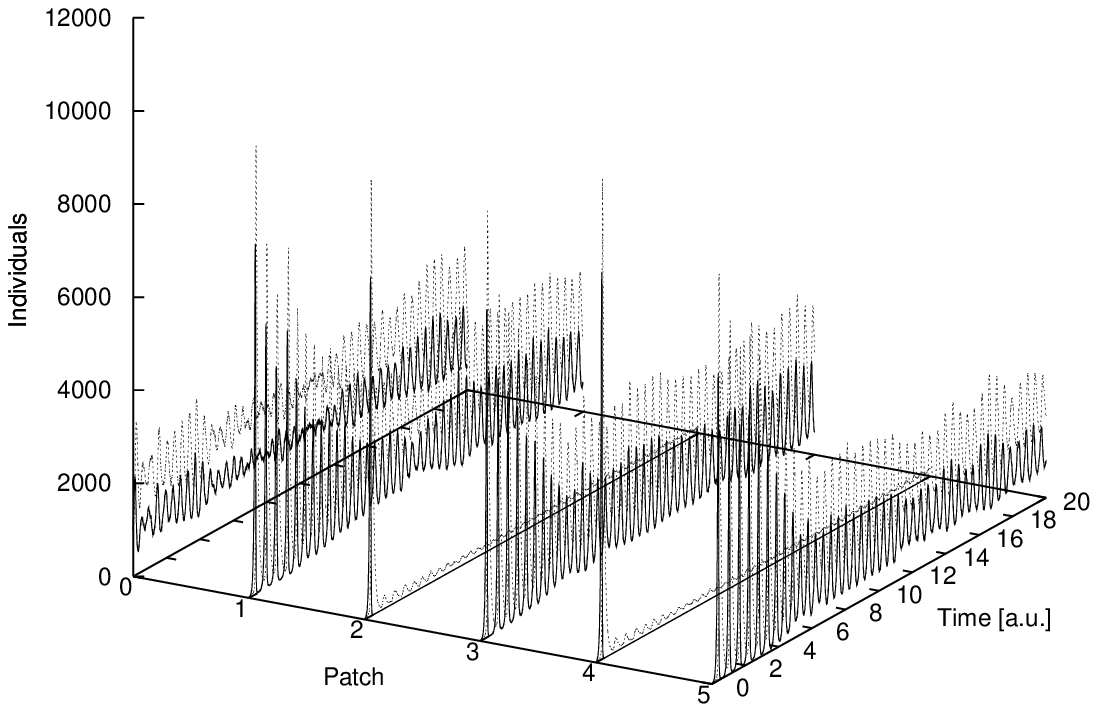} \hspace{-0.7cm}
\includegraphics[width=7.5cm]{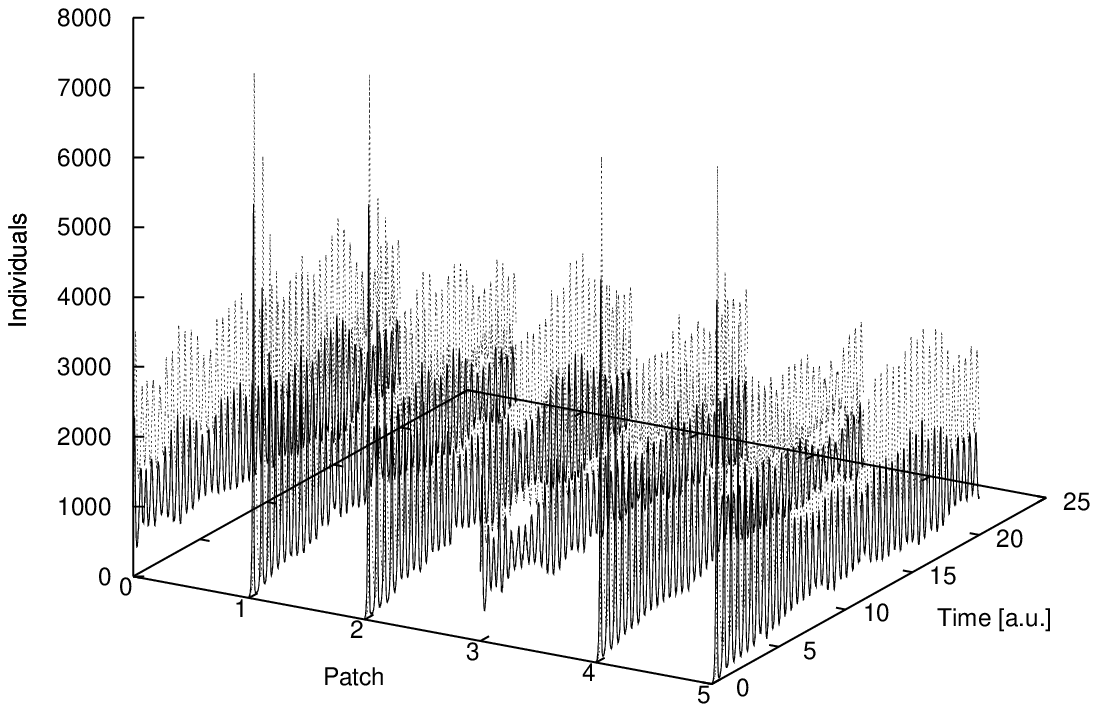}
\end{center}
\caption{Colonization in the complete topology, with initial condition IC1 and $p_{LV}^{e}$=$\{p_0\}$ (left), $p_{LV}^{e}$=$\{p_0, p_3\}$ (right).} \label{fig:complete}
\end{figure}

For the network topology $\nu=e$, that is, the complete graph, we show the results obtained in the cases IC$1$, when $p_{LV}^{e}$=$\{p_0\}$ (Fig. \ref{fig:complete}, left side) and $p_{LV}^{e}$=$\{p_0, p_3\}$ (Fig. \ref{fig:complete}, right side). While in the second case -- where the LV dynamics is initially placed in two patches -- the predators can colonize all patches, in the first case the colonization of all empty patches fails. Once more, this is an effect of the stochastic noise combined with the low amounts of predators, which is in turn caused by the fact that the higher the number of adjacent patches, the lower the number of predators that persist inside each patch. In all other simulations performed with initial conditions IC3 and IC4, all patches have always been colonized, as the higher values of dispersal rules assure a more uniform spread of predators throughout the network, and thus flattens the influence of migration delay (data not shown).

\begin{figure}[htbp]
\begin{center}
\includegraphics[width=7.5cm]{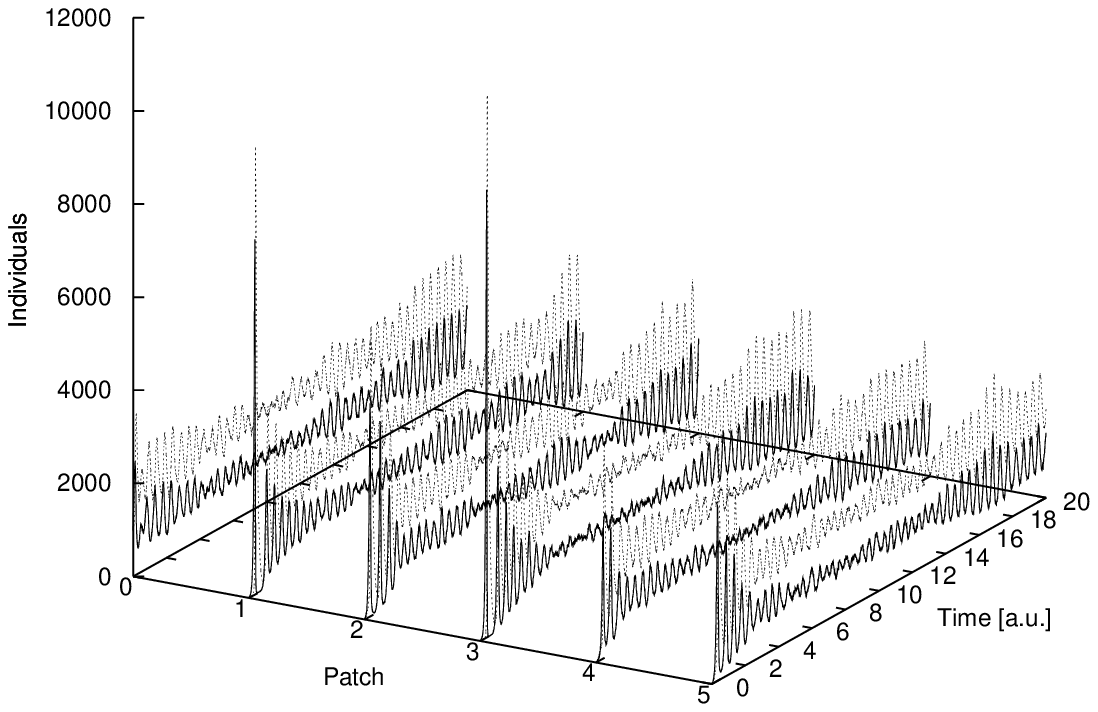} \hspace{-0.7cm}
\includegraphics[width=7.5cm]{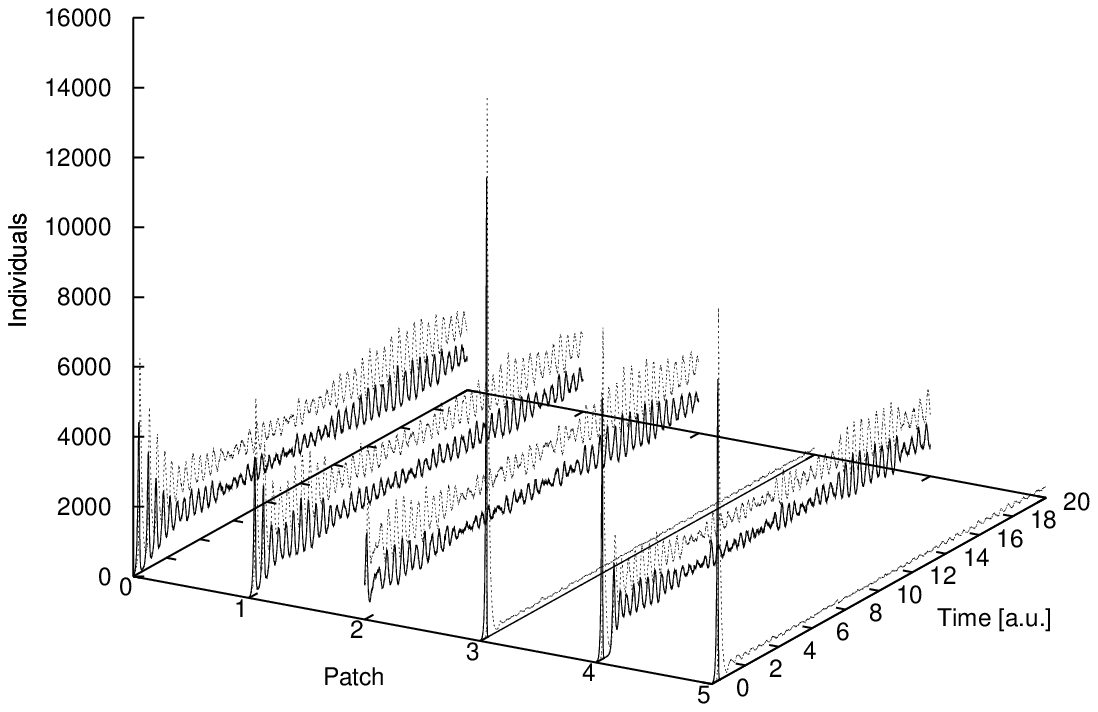}
\end{center}
\caption{Colonization in the random topology, with initial condition IC1 and $p_{LV}^{f}$=$\{p_0\}$ (left), $p_{LV}^{f}$=$\{p_2\}$ (right).} \label{fig:random}
\end{figure}

For the network topology $\nu=f$, that is, the random graph, we show the results obtained in the cases IC$1$, when $p_{LV}^{f}$=$\{p_0\}$ (Fig. \ref{fig:random}, left side) and $p_{LV}^{f}$=$\{p_2\}$ (Fig. \ref{fig:random}, right side). According to the position of the LV complete patches in this network topology, we can see that, in the first case, all patches are colonized by predators (similar results are obtained by placing the LV complete model in patch $p_4$ -- data not shown). In the second case, patch $p_5$ is not colonized because there is only one path of length 2 which connects it to the initial complete LV patch $p_2$; the same holds for patch $p_3$, which has distance from $p_2$ equal to 3. For similar reasons, considering the case of initial condition IC1, with the  LV complete model in patch $p_3$, the only patch that is not colonized by predators is $p_2$ (data not shown).
In all the simulations performed with the initial condition IC2, some of the patches have not been colonized because of the high amount of preys initially occurring in the patches. On the other hand, with the initial conditions IC3, IC4, the power of dispersal allows the colonization of all patches (data not shown).

%%%%%%%%%%%%%%%%%%%%%%%%%%%%%%%%%%%%%%%%%%%%%%%%%%%%%%%%%
%%%%%%%%%%%%%%%%%%%%%%%%%%%%%%%%%%%%%%%%%%%%%%%%%%%%%%%%%

%%%%%%%%%%%%%%%%%%%%%%%%%%%%%%%%%%%%%%%%%%%%%%%%%%%%%%%%%
%%%%%%%%%%%%%%%%%%%%%%%%%%%%%%%%%%%%%%%%%%%%%%%%%%%%%%%%%

\section{Discussion}\label{sec:concl}

The fragmented habitats of real metapopulations are usually characterized by complex network topologies. In this paper, we have analyzed six small topologies that can be considered representative of local areas in a structured habitat, and we have investigated the influence that the degree and the position of each patch in the topology can have on the migration of individuals, as well as on the capability of colonizing empty patches. Our analysis suggests that, with respect to the power of migration (Section \ref{subsec:communication}), we can identify different behaviours that depend on two characteristics of the topology: on a first level, the local behaviour inside each patch is influenced by its degree. This is especially evident if we compare the network topology described by the circular or complete graphs, with the topology described by the star graph: while in the first case (where all nodes have the same degree) all patches are characterized by a similar (regular) oscillating dynamics, in the second case the most critical node is the center of the star (which has a much higher degree than all other nodes in the same graph). In the latter case, this patch is likely to undergo a local modification of its initial dynamics, due to a more higher incoming migration of individuals from all other adjacent patches. On a second level, assuming in this case that the degree of nodes is equal, then also the position of each patch in the topology matters: for instance, we have seen that in the network topology described by the chain graph -- where all nodes, besides the ones at the extremes of the chain, have the same degree -- the local dynamics is also influenced by the dynamics of the adjacent patches in the graph. Therefore, in hypothetical habitats where there exist many patches connected in a linear way, our results suggest that the length of the chain might have a negative role in the establishment and in the maintenance of local dynamics.

Considering the feature of colonization (Section \ref{subsec:colonization}), we have evidenced that, in most network topologies, the lack of colonization can be due to the delay of migrating predators with respect to the (uncontrolled) local growth of prey, which then leads to the extinction of preys and the prevention of the LV dynamics. To effectively measure how strong is the power of the delay, it would be interesting to understand whether the local growth of preys can be controlled by inducing their death and thus potentially allowing the establishment of oscillations. Besides this aspect deserving further investigations, our analysis have evidenced that the colonization of empty patches occurs more easily in those patches that are adjacent to the patch(es) initialized with the LV complete model. Once more, this highlights the relevance of the position of the patch(es) where standard oscillations in preys and predators are already settled at the beginning of the simulation. Indeed, the power of colonization is stronger in the circular and complete networks -- where the position of the LV complete patch is irrelevant (as the spread of migrating individuals throughout the network results uniform), and it is weaker in the star network -- where the position of the LV complete patch is of primary importance (as the spread of migrating individuals throughout the network strongly depends on whether the patch is placed at the center or at the tips of the star).

In addition to the investigations that we have presented in this work, further types of analysis that we plan to perform on metapopulation systems concern, for instance, the study of the aspects considered in this paper (migration, colonization, network topologies, etc.) by assuming other local and global dynamics, e.g., the population growth according to the logistic function. Moreover, an interesting issue that might be investigated is the synchronization of local population dynamics (e.g. by considering the establishment and decay of oscillations in preys and predators) during migration through a given network topology, or in the process of colonization.

Concerning the use of graphs, other relevant questions regard the analysis of the dynamics with respect to graph properties, such as different measures of habitat connectivity (centrality indexes) \cite{evolution_networks,newman_siam}. In this context, for example, the star graph can resemble the notion of hub (a node with high degree) in a typical scale-free network, a structure that is known to be robust to random disturbances but highly vulnerable to deliberate attacks on the hubs \cite{strogatz,barabasi}.

Another topic of interest concerns the fact that various populations can coexist in a common habitat, but have distinct (inter)species dynamics or different dispersal capabilities in that habitat \cite{Bunn00}. In cases like this, it would be interesting to construct and analyze different metapopulation models, one for each target species, according to both the patch-to-patch connections and to the specific population dynamics. By comparing and intersecting the results obtained on the distinct network topologies of the common habitat derived in this way, it would be possible to determine the locations of the habitat that are most important for each species, and thus aid the design of natural reserve systems where we can have the most appropriate solution for all species in terms of the maximal improvement of dispersal (reduction of species isolation) and the minimal spread of disturbances (diseases, pathogens, invasive species, etc.) \cite{habitat_mosaics}.

We believe that our modeling approach opens interesting perspectives and can represent an useful tool for the investigation of a wide range of properties in metapopulation systems. We expect that applications of this model to real cases -- characterized by complex habitat networks (where each patch possesses its own features of quality, occupancy, connectivity) and different population dynamics -- will aid in the achievement of important results and new perspective in Ecology.

\bibliographystyle{eptcs}
\bibliography{bibliography}

\end{document}